\documentclass[preprintnumbers,prd,nofootinbib,floats,amssymb,floatfix]{revtex4-2}
\usepackage{amsfonts}
\usepackage{amsmath}
\usepackage{graphicx}
\usepackage{color}
\usepackage{amssymb}
\usepackage{hyperref}
\hypersetup{colorlinks,linkcolor={cyan},citecolor={cyan},urlcolor={magenta}}
\usepackage[normalem]{ulem}

\begin{document}

\title{Canonical description of  Exotic General Massive Gravity}

\author{Omar Rodr\'iguez-Tzompantzi}
\email{omar.tzompantzi@unison.mx}
\affiliation{Departamento de Investigaci\'on en F\'isica, Universidad de Sonora,  Apartado Postal 5-088, C.P. 83000, Hermosillo, Sonora, M\'exico}

\begin{abstract}
Exotic General Massive Gravity is the next-to-simplest gravitational theory fulfilling the so-called third-way consistency, the simplest being Minimal Massive Gravity. We investigate the canonical structure of the first-order formulation of Exotic General Massive Gravity. By using the Dirac Hamiltonian formalism, we systematically discover the complete set of physical constraints, including primary, secondary, and tertiary ones, and explicitly compute the Poisson bracket algebra between them. In particular,  we demonstrate that the consistency condition for the tertiary constraints provides explicit expressions which can be solved algebraically for the auxiliary fields $f$ and $h$ in terms of the dreibein $e$. In this configuration, to confirm that the theory is ghost-free, the whole set of constraints is classified into first and second-class ones showing the existence of only two physical degrees of freedom corresponding to  one massive graviton. Furthermore, we identify the transformation laws for all of the dynamical variables corresponding essentially to gauge symmetries, generated by the first-class constraints. Finally, by taking into account all the second-class constraints, we explicitly compute the Dirac matrix together with the Dirac's brackets.
\end{abstract}

\maketitle

\section{Introduction}
\label{introduction}
Einstein's theory of General Relativity (GR) is accepted as the unique four-dimensional theory of the gravitational interaction \cite{Einstein,Einstein2,Lovelock,Lovelock2,Dyson}, that predicts the propagation of only two physical degrees of freedom (DoF), corresponding to a single massless spin-two field - the graviton. The consistency of these interactions is guaranteed by the invariance of the theory under spacetime diffeomorphism symmetry which also provides GR with a unique and intuitive geometrical interpretation.  Furthermore, several predictions of GR have been observationally confirmed. For instance,  the precession of Mercury's orbit \cite{Perihelion}, gravitational lensing \cite{Lens,Lens2}, gravitational waves \cite{LIGO}, and most recently black holes \cite{Schwarzschild,Horizon}. Notwithstanding these positive results, pure GR is not well suited for quantization, since the resulting theory of quantum gravity proves to be non-renormalizable. However, particular implementations of higher-derivative terms to the Einstein-Hilbert action, such as  Ricci-squared and scalar curvature-squared terms, make the theory renormalizable at the cost of the loss of unitary \cite{Hooft,Stelle}. This fact has led many theorists to search for extensions of GR and their three-dimensional (3D) counterparts as suitable candidates to give hints to the solution of the real problems in four dimensions. This  because they provide a powerful background to elucidate conceptual problems and look for quantum aspects of gravity, while keeping the technical problems at a bare minimum.

Gravity in three dimensions,  with or without a cosmological constant,  is a topological field theory in the sense that it does not possess local DoF, and therefore there is no propagation of gravitons \cite{Deser,Witten}. Nonetheless, it allows non-perturbative quantizations following different approaches \cite{LGQ,LGQ2,Porrati,Nelson,Riello}, which makes 3D gravity a nice toy model of quantum gravity. Interestingly enough, in spite of having no DoF, 3D gravity can be generalized to propagate physical DoF, which makes the theory closer to its four-dimensional cousin; it can be attained by adding certain higher-derivative terms to gravity action. One of the most important proposals in this regard is Topologically Massive Gravity (TMG) \cite{Deser1, Deser2} which consists of the usual Einstein-Hilbert term, a negative cosmological constant, and one parity-violating gravitational Chern-Simons term. Remarkably, the TMG model is unitary and leads to third-order field equations, whose linearization about an Anti-de Sitter (AdS) vacuum, yields a single massive spin-2 mode. Besides that, the theory allows for black hole solutions which have AdS asymptotics \cite{Bouchareb, Hotta}. 

A remarkable extension of TMG is New Massive Gravity (NMG) \cite{NMG} which is parity-even and quadratic in the curvature, such that the field equations for the metric are of fourth-order. At the linearized level, this higher-derivative theory is equivalent to the 3D version of the Pauli-Fierz theory \cite{Fierz,HigherDerivatives} for a massive spin-2 field. If one abandons the parity-preserving condition on NMG, the theory can be extended to General Massive Gravity (GMG) \cite{NMG} that propagates two massive gravitons with different masses, and that also includes TMG as one special case. It is worthwhile noting that all the theories of 3D massive gravity discussed up to this point arise from diffeomorphism invariant actions; hence the resulting field equations are divergence-free. That is to say, the rank two tensor defining the field equations satisfies a Bianchi identity for all smooth metrics.

Interestingly, other than TMG, NMG, and GMG, there is a new class of 3D massive gravity theories, labelled as third-way consistent, which have field equations that contain tensors whose covariant divergences vanish only on-shell. These models do not have actions that merely contain the metric field, and it is not only the Bianchi identity  that guarantees the consistency of the equations of motion \cite{Bergshoeff,Matteo}. The simplest such example is the so-called Minimal Massive Gravity (MMG) \cite{Bergshoeff2, Arvanitakis}, which is defined by supplementing the equations of motion of TMG with a rank-2 tensor containing up to second derivatives of the metric. This MMG theory is unitary within a certain range of its parameters \cite{Arvanitakis} and has the same physical DoF as the TMG, such that the theory involves a single propagating graviton as well. The next-to-simplest example is Exotic Massive Gravity (EMG) \cite{EMG}.   This model can be attained by supplementing the Einstein equations with a term that contains up to third derivatives of the metric and is constructed with combinations and derivatives of the Cotton tensor, which can also be added by itself with its own coupling, while still maintaining the consistency of the field equations; this in turn propagates a parity-doublet of massive spin-2 modes. In this paper, we will focus on a slight modification of EMG, dubbed as Exotic General Massive Gravity (EGMG).

The EGMG model is a parity-violating generalization of EMG; it propagates a parity doublet of massive spin-2 modes but with the mass degeneracy lifted by parity-violation \cite{EMG,EGMG}. Although there is no diffeomorphism invariant action whose variation yields the field equation of the theory in the metric formulation \cite{Bergshoeff}, a Lagrangian formulation of the theory can be constructed in the first-order formalism by considering the dreibein, the dual spin-connection, and two auxiliary fields. Such auxiliary fields are very important if we want a gravity model with a higher number of derivatives in the metric formulation. In particular, its first-order formulation has been useful in computing  the central charge of the boundary theory \cite{Merbis}. 

So far, the canonical description and constraint analysis of EGMG theory has not been investigated. Understanding the nature of the constraints and their effects on the canonical description of higher-order massive gravity models is an issue of prime importance in both classical and quantum gravity.  Strictly speaking, the Hamiltonian description of any classical physical theory consists of a Hamiltonian functional and a phase-space equipped with a symplectic structure. The phase-space is a specification of the physical DoF in the theory, while the Hamiltonian functional describes how these DoF evolve in time. The split between the phase-space and the Hamiltonian functional is also relevant when quantizing the system, as one must replace the phase-space by a Hilbert space, while the Hamiltonian functional is replaced by a Hermitian operator.  Moreover, all theories of the fundamental interactions are characterized by the presence of gauge symmetry which is related, from a Hamiltonian point of view, to the existence of constraints on the phase-space variables  and on the dynamics of the theory. Indeed, such constraints are precisely the generators of the gauge transformations. Notably, for gravity theories, this symmetry (being the canonical realization of the general principle of relativity)  not only represents the invariance of the physical content of the theory concerning changes of coordinates, but is also intricately related to the dynamics of the theory \cite{Bodendorfer}. We recall the fact that, from a canonical point of view, GR is a theory fully governed by constraints\cite{Dirac,Henneaux,P. A. M,Blagojevic, Ashtekar}, meaning that the Hamiltonian functional is a linear combination of constraints,  and consequently, its dynamics is completely dictated by constraints. Hence, these constraints are expected to be crucial at the quantum level. This remarkable feature suggests that massive gravity theories must possess the necessary physical constraints to describe the dynamical structure of these types of theories, however, the identification of these constraints is notoriously difficult, and the model considered in this paper is not an exception. Motivated by its possible quantization and by reaching a further understanding of the EGMG itself, a complete and consistent canonical analysis of EGMG turns out to be of great importance.

The main aim of this paper is to provide a detailed study of the dynamical content of EGMG non-perturbatively, by using the Dirac Hamiltonian approach. According Dirac's procedure, the first main step is to extract the primary constraints from the definition of the momenta. The basic consistency of the theory requires that the primary constraints be conserved during the dynamical evolution of the system. Then these conditions of consistency will generate the secondary constraints, and so on. Thereafter, the complete set of constraints must be classified into first and second-class ones. When it does, the number of physical DoF can be explicitly counted independently of any linearized approximation, and a generator of the gauge transformations can be constructed as a suitable combination of the first-class constraints.  Furthermore, Dirac's bracket structure which is important for quantizing the model can be obtained once the second-class constraints have been eliminated as strong equalities. All of these elements are potentially useful for one possible Dirac-type quantization in the  canonical framework of loop quantum gravity \cite{LGQ,LGQ2}, for example. In this paper, we investigate and classify, in a rather systematic manner, the different constraints according to Dirac terminology, as detailed below.

The paper is organized as follows. In the next section, \ref{EGMG}, we will introduce the action principle corresponding to EGMG. Later, after introducing the action principle, we will perform the $2 + 1 $ decomposition of the action in section \ref{Constraints}, to perform the full Hamiltonian
analysis. We will then show that a careful analysis of the dynamical structure reveals the complete set of physical constraints, including primary, secondary and tertiary. In section \ref{classify}, we deal with finding the first and second-class constraints in EGMG theory and identifying the physical degrees of freedom associated with our model. Then we determine the fundamental quantization brackets and recover the symmetry of diffeomorphisms, by using the first and second-class constraints respectively. Finally, we establish our main conclusions in section \ref{conclusions}, and some of the
calculational details are relegated to appendices \ref{Algebra} and \ref{DB} .

\section{The action}
\label{EGMG}
We consider the following first-order action for Exotic General  Massive Gravity \cite{EMG,EGMG}:
\begin{eqnarray}
S\left[w,e,f,h\right]&=&\int\limits_{\mathcal{M}}\,\left[f^{I}\wedge R_{I}+\frac{1}{6m^{4}}\epsilon^{IJK}f_{I}\wedge f_{J}\wedge  f_{K}-\frac{1}{2m^{2}}f^{I}\wedge \mathrm{D}f_{I}+\frac{\nu}{2}\epsilon^{IJK}f_{I}\wedge e_{J}\wedge  e_{K}-m^{2}h^{I}\wedge \mathrm{T}_{I}\right.\nonumber\\
&&\left.\frac{1}{2}\left(\nu-m^{2}\right)w^{I}\wedge\left(d w_{I}+\frac{1}{3}\epsilon_{IJK} w^{J}\wedge  w^{K}\right)+\frac{1}{3}\frac{\nu m^{4}}{\mu}\epsilon^{IJK}e_{I}\wedge e_{J}\wedge e_{K}\right],\label{action}
\end{eqnarray}
with $\nu=\left(1/l^{2}-m^{4}/\mu^{2}\right)$;  where $m$ and $\mu$ are mass parameters, and $l$ is the AdS radius of curvature. The fundamental fields of this action are: the dreibein 1-form $e^{I}=e^{I}_{\mu}dx^{\mu}$ that determines a space-time metric via $g_{\mu\nu}=e^{I}_{\mu}e^{J}_{\nu}\eta_{IJ}$; a pair of auxiliary fields 1-form $f$ and $h$; and the dualized spin-connection $w^{I}=w^{I}_{\mu}dx^{\mu}$ valued on the
adjoint representation of the Lie group $SO(1,2)$, so that, it admits a totally invariant anti-symmetric tensor $\epsilon^{IJK}$.  Furthermore, $\mathrm{T}_{I}$ and  $R^I$ are the torsion and curvature two-forms, respectively: 
\begin{eqnarray}
\mathrm{T}_{I}&=&d e_{I}+\epsilon^{I}{_{JK}}w^{J}\wedge e^{K},\\
R^{I}&=&dw^{I}+(1/2)\epsilon^{I}{_{JK}}w^{J}\wedge w^{K}.
\end{eqnarray}
 In particular, we defined the standard covariant derivative, acting on internal indices, by means of the spin-connection:
\begin{equation}
D_{\mu}V^{I}=\partial_{\mu}V^{I}+\epsilon^{I}{_{JK}}w_{\mu}^{J}V^{K},\quad\text{with}\quad V^{I}\in(e^{I},f^{I},h^{I}),
\end{equation}
where $\partial$ is a fiducial derivative operator. The convention adopted is the standard one, that is $x^\mu$, $\mu = 0, 1, 2$, are local coordinates that label the points of the three-dimensional oriented smooth manifold $\mathcal{M}$. Whereas, the Latin capital letters $I$ correspond to Lorentz
indices, $I = 1, 2, 3$. 

In the metric formulation, EGMG is defined by the following equations of motion:
\begin{equation}
R_{\mu\nu}-\frac{1}{2}g_{\mu\nu}R+\Lambda g_{\mu\nu}+\frac{1}{\mu}C_{\mu\nu}-\frac{1}{m^{2}}H_{\mu\nu}+\frac{1}{m^{4}}L_{\mu\nu}=0.
\end{equation}
With
\begin{eqnarray}
H_{\mu\nu}&=&\varepsilon_{\mu}{^{\alpha\beta}}\nabla_{\alpha}C_{\nu\beta},\\
L_{\mu\nu}&=&\frac{1}{2}\varepsilon_{\mu}{^{\alpha\beta}}\varepsilon_{\nu}{^{\gamma\rho}}C_{\alpha\gamma}C_{\beta\rho},\\
C_{\mu\nu}&=&\varepsilon_{\mu}{^{\alpha\beta}}\nabla_{\alpha}\left(R_{\mu\nu}-\frac{1}{4}g_{\mu\nu}R\right),
\end{eqnarray}
here $H_{\mu\nu}$ and $L_{\mu\nu}$ are traceless and symmetric tensors, $C_{\mu\nu}$ is the Cotton tensor which is symmetric, traceless, and divergence-free, and $\Lambda=-1/l$ is the cosmological constant. These two formalisms are equivalent, at least at the level of equations of motion, assuming the invertibility of the dreibein. Linearization about an AdS vacuum yields two massive spin-2 excitations with different masses, given as \cite{EMG}\begin{equation}
M_{\pm}= m\left[\sqrt{1+\frac{m^{2}}{4\mu^{2}}}\pm\frac{m}{2\mu}\right],
\end{equation}
which gives $M_{\pm} = m$ in the EMG ($\mu\rightarrow\infty$) limit. This result for $M_{\pm}$ is independent of $l$ and therefore applies in the Minkowski limit, for which $M_{\pm}$ are the masses of the two propagating modes, of helicities $\pm 2$.

\section{The nature of constraints and its stability}
\label{Constraints}
To carry out Hamiltonian analysis, not only do we assume that the spacetime $\mathcal{M}$ is globally hyperbolic such that it may be foliated as $\Sigma\times\Re$, with $\Sigma$ being a Cauchy's surface without boundary and $\Re$ an evolution parameter, but also that simultaneous proper $2+1$ decompositions exist for the dreibein $e$ and for the pair of auxiliary fields $f$, $h$.  By performing the $2 + 1$ decomposition of our fields, up to surface terms, the action (\ref{action}) can be written in the form
\begin{equation}
S=\int\limits_{\Sigma\times\Re}dtd^{2}x\left[\varepsilon^{0ab}\dot{w}_{a}^{I}\left(f_{b I}+\frac{1}{2}\left(\nu -m^{2}\right)w_{bI}\right)-\frac{1}{2m^{2}}\varepsilon^{0ab}\dot{f}_{a}^{I}f_{b I}-m^{2}\varepsilon^{0ab}\dot{e}_{a}^{I}h_{bI}+f_{0}^{I}\Phi_{I}+w_{0}^{I}\Psi_{I}+e_{0}^{I}\Sigma_{I}-h_{0}^{I}\Theta_{I}\right].\label{2+1}
\end{equation}
In the action above, the dot stands for derivation with respect to time, and the twelve functions $\Phi_{I}$, $\Psi_{I}$, $\Sigma_{I}$, $\Theta_{I}$ have the following form:
\begin{eqnarray}
\Phi_{I}&=&\varepsilon^{0ab}\left(R_{ab}^{I}+\frac{1}{2m^{4}}\epsilon^{IJK}f_{a J}f_{b K}-\frac{1}{m^{2}}\mathrm{D}_{a}f_{b}^{I}+\frac{\nu}{2}\epsilon^{IJK}e_{a J}e_{b K}\right)\label{s1},\\
\Psi_{I}&=&\varepsilon^{0ab}\left(\mathrm{D}_{a}f_{b}^{I}-\frac{1}{2m^{2}}\epsilon^{IJK}f_{a J}f_{b K}-m^{2}\epsilon^{IJK}h_{a J}e_{b K}+\left(\nu-m^{2}\right)R_{ab}^{I}\right)\label{s2},\\
\Sigma_{I}&=&\varepsilon^{0ab}\left(\nu\epsilon^{IJK}f_{a J}e_{b K}-m^{2}\mathrm{D}_{a}h_{b}^{I}+\frac{\nu m^{4}}{\mu}\epsilon^{IJK}e_{a J}e_{b K}\right)\label{s3},\\
\Theta_{I}&=&\varepsilon^{0ab}m^{2}\mathrm{D}_{a}e_{b}^{I}\label{s4},
\end{eqnarray}
where the spatial components of the curvature and the covariant derivative have the following structure:
\begin{eqnarray}
R_{ab}^{I}&=&\partial_{a}w_{b}^{I}+\frac{1}{2}\epsilon^{I}_{JK}w_{a}^{J}w_{b}^{K},\\ \mathrm{D}_{a}V^{I}_{b}&=&\partial_{a}V_{b}^{I}+\epsilon^{I}_{JK}w_{a}^{J}V^{K}_{b}.
\end{eqnarray}
Before we discuss the Hamiltonian analysis, it is important to appreciate here that the action (\ref{2+1}) describes the evolution of $36$ coordinates fields: $f^{I}_{\mu}$ ($3\times3=9$), $w^{I}_{\mu}$ ($3\times3=9$), $e^{I}_{\mu}$ ($3\times3=9$) and $h^{I}_{\mu}$ ($3\times3=9$); among them there are $12$ ($f^{I}_{0}$, $w^{I}_{0}$, $e^{I}_{0}$, $h^{I}_{0}$) whose time derivatives do not appear in the action (\ref{2+1}), in such a way that they can be considered as Lagrange multipliers for phase-space constraints from the beginning. Throughout this work, we will not associate conjugate momenta to these variables in order not to unnecessarily complicate the Hamiltonian analysis. Indeed, it is straightforward to check that the variation of the action with respect to Lagrangian multipliers yields the following conditions:
\begin{equation}
\Phi_{I}=0,\quad\Psi_{I}=0,\quad \Sigma_{I}=0,\quad \Theta_{I}=0.\label{Equations2}
\end{equation}
We should take these as the Lagrangian constraints. 

On the other hand, it is simple to see that  the action (\ref{2+1}) is first-order in time derivative of  the variables  $w^{I}_{a}, f^{I}_{a}, e^{I}_{a},  h^{I}_{a}$, and therefore the action can be compactly written as
\begin{equation}
S=\int\limits_{\Sigma\times\Re}dtd^{2} x\left(a_{i}[\xi]\dot{\xi}^{i}-\mathcal{V}[\xi]\right).\label{first-order}
\end{equation}
The form of $S$ given above is such that  $\xi^{i}=(w^{I}_{a}, f^{I}_{a}, e^{I}_{a},  h^{I}_{a})$ stands for the collection of all the dynamical fields of the theory,  $a_{i}=(\varepsilon^{0ab}\left(f_{b I}+\frac{1}{2}\left(\nu -m^{2}\right)w_{bI}\right),-\frac{1}{2m^{2}}\varepsilon^{0ab}f_{b I},-m^{2}\varepsilon^{0ab}h_{bI},0)$ are the components of the so-called canonical one-form $a=a_{i}d\xi^{i}$  as defined in Ref. \cite{F-J}, and $\mathcal{V}=f_{0}^{I}\Phi_{I}+w_{0}^{I}\Psi_{I}+e_{0}^{I}\Sigma_{I}-h_{0}^{I}\Theta_{I}$ represents one potential density which could also be identified with the canonical Hamiltonian density.  As a consequence, the equations of motion follow by extremizing the action (\ref{first-order})  with respect to the dynamical variables $\xi^{i}$. They are given by
\begin{equation}
\int\limits_{\Sigma}d^{2} x\left(\mathcal{F}_{ij}\dot{\xi}^{j}-\frac{\delta\mathcal{V}}{\delta\xi^{i}}\right)=0,\label{AnotherEM}
\end{equation}
where $\mathcal{F}_{ij}=\delta a_{i}/\delta\xi^{j}-\delta a_{j}/\delta\xi^{i}$ is  the called pre-symplectic two-form matrix which is defined as a generalized curl of the canonical one-form. For our purposes in the present paper, it is not necessary to specify the explicit form of the matrix $\mathcal{F}_{ij}$. However, it is important to notice that $\mathcal{F}_{ij}$ can be either singular or non-singular. On the one hand, if the matrix $\mathcal{F}_{ij}$ is non-singular, then its inverse can be computed. As a consequence, the symplectic structure and the set of equations of motion  could immediately be derived following the formalism of \cite{F-J}. On the other hand, if the matrix $\mathcal{F}_{ij}$ is singular, then there are more degrees of freedom in the equations of motion than physical degrees of freedom in the theory. Thus, there must be constraints  to maintain the consistency of the equations of motion.

 To proceed with the Hamiltonian formulation of EGMG we shall use the Dirac formalism \cite{Dirac}. The first step is to define  momenta canonically conjugate to the dynamical variables. Since $S=\int_{\Re} dt\int_{\Sigma} d^{2}x \mathcal{L}$ (\ref{2+1}),  the corresponding conjugated momenta are defined  as,
 \begin{eqnarray}
\Pi^{a}_{I}=\frac{\partial{\mathcal{L}}}{\partial\dot{f}_{a}^{I}},\qquad\pi^{a}_{I}=\frac{\partial{\mathcal{L}}}{\partial\dot{w}_{a}^{I}},\qquad{\widetilde{\Pi}}^{a}_{I}=\frac{\partial{\mathcal{L}}}{\partial\dot{e}_{a}^{I}},\qquad{\widetilde{\pi}}^{a}_{I}=\frac{\partial{\mathcal{L}}}{\partial\dot{h}_{a}^{I}}.
\label{eq:p11}
\end{eqnarray}
From (\ref{eq:p11}), we immediately get to the following primary constraints:
\begin{eqnarray}
\phi^{a}_{I}&=& \Pi^{a}_{I}+\frac{1}{2m^{2}}\varepsilon^{0ab}f_{b I}\approx0,\label{primary1} \\
\psi^{a}_{I}&=&\pi^{a}_{I}-\varepsilon^{0ab}\left(f_{bI}+\frac{1}{2}\left(\nu -m^{2}\right)w_{bI}\right)\approx0,\label{primary2} \\
\sigma^{a}_{I}&=&\widetilde{\Pi}^{a}_{I}+m^{2}\varepsilon^{0ab}h_{bI}\approx0,\label{primary3}\\
\theta^{a}_{I}&=& \widetilde{\pi}^{a}{_{I}}\approx0\label{primary4},\label{primary}
\end{eqnarray}
where $\approx$ stands for weak equality in the sense of Dirac \cite{Dirac}, implying that it is only valid on the so-called primary constraint surface $\Gamma_{P}\subset\Gamma$ defined by the primary constraints and contained in the phase-space. The phase-space $\Gamma$ of this model includes
the dynamical fields $w^{I}_{a}$, $f^{I}_{a}$, $e^{I}_{a}$, $h^{I}_{a}$, and their conjugate momenta $\pi^{a}_{I}$, $\Pi^{a}_{I}$, $\widetilde{\Pi}^{a}_{I}$, $\widetilde{\pi}^{a}_{I}$. Note also that,  as stated above, the twelve expressions $\Phi_{I},\Psi_{I},\Sigma_{I},\Theta_{I}$ defined in Eqs. (\ref{s1})-(\ref{s4}) are secondary constraints associated with the Lagrange multipliers $f^{I}_{0}$, $w^{I}_{0}$, $e^{I}_{0}$ and $h^{I}_{0}$, respectively. In this way, the canonical Hamiltonian weakly vanishes and therefore the dynamics is generated by the so-called primary Hamiltonian which is a linear combination of all the above constraints,
\begin{equation}
H_{\text{P}}=\int d^{2}x\left[\lambda_{a}^{I}\phi^{a}_{I}+\upsilon^{a}_{I}\psi_{a}^{I}+\vartheta^{a}_{I}\sigma_{a}^{I}+\iota^{a}_{I}\theta_{a}^{I}+h_{0}^{I}\Theta_{I}-f_{0}^{I}\Phi_{I}-w_{0}^{I}\Psi_{I}-e_{0}^{I}\Sigma_{I}\right],\label{HamiltonPrimery}
\end{equation}
where $\lambda^{I}_{a}, \upsilon^{I}_{a},\vartheta^{I}_{a},\iota^{I}_{a}$, play the role of the Lagrange multipliers enforcing the primary constraints. In this theory, the non-vanishing fundamental Poisson brackets are given by
\begin{equation}
\{f_{a}^{I}(\mathbf{x}),\Pi_{J}^{b}(\mathbf{y})\}=\{w_{a}^{I}(\mathbf{x}),\pi_{J}^{b}(\mathbf{y})\}=\{e_{a}^{I}(\mathbf{x}),\widetilde{\Pi}_{J}^{b}(\mathbf{y})\}=\{h_{a}^{I}(\mathbf{x}),\widetilde{\pi}_{J}^{b}(\mathbf{y})\}=\delta_{a}^{b}\delta_{J}^{I}\delta^{2}(\mathbf{x}-\mathbf{y}),\label{Poisson1}
\end{equation}
where $\mathbf{x}$ and $\mathbf{y}$ stand for the spatial coordinates  $x^{a}$ and $y^{a}$ of the spatial slice $\Sigma$, while $\delta^{2}(\mathbf{x}-\mathbf{y})$ is the Dirac delta function in two dimensions.  Generally, any Poisson bracket involving arbitrary functionals of fields and conjugate momenta can be computed using these fundamental relations. Under this observation, the Poisson bracket of any functional $F$ of the canonical variables with the Hamiltonian provides its time evolution, namely,
\begin{equation}
\frac{d}{dt}F=\dot{F}=\{F,H_{\text{P}}\}.\label{evolution}
\end{equation}
A basic consistency\footnote{These consistency conditions will either solve some multipliers, or lead to the new constraints, or will be identically satisfied.} requirement on the theory is that the primary constraints  be preserved in time,  which guarantees that such constraints remain on the constraint surface $\Gamma_{\text{P}}$ during their evolution. Preservation of primary constraints can be written as
\begin{equation}
\dot{\phi}^{a}_{I}=\{\phi^{a}_{I},H_{\text{P}}\}\approx0, \quad\dot{\psi}^{a}_{I}=\{\psi^{a}_{I},H_{\text{P}}\}\approx0, \quad\dot{\sigma}^{a}_{I}=\{\sigma^{a}_{I},H_{\text{P}}\}\approx0, \quad\dot{\theta}^{a}_{I}=\{\theta^{a}_{I},H_{\text{P}}\}\approx0.\label{ConsistencyPrimary}
\end{equation}
 If we want to compute these consistency conditions,  we need the non-vanishing Poisson brackets of all the primary constraints among themselves,
\begin{eqnarray}
\{\phi_{I}^{a}(\mathbf{x}),\phi_{J}^{b}(\mathbf{y})\}&=&\frac{1}{m^{2}}\varepsilon^{0ab}\eta_{IJ}\delta^{2} (\mathbf{x}-\mathbf{y}),\label{pri1}\\
\{\phi^{a}_{I}(\mathbf{x}),\psi^{b}_{J}(\mathbf{y})\}&=&-\varepsilon^{0ab}\eta_{IJ}\delta^{2} (\mathbf{x}-\mathbf{y}),\label{pri2}\\
\{\psi^{a}_{I}(\mathbf{x}),\psi^{b}_{J}(\mathbf{y})\}&=&-\varepsilon^{0ab}\left(\nu-m^{2}\right)\eta_{IJ}\delta^{2} (\mathbf{x}-\mathbf{y}),\label{pri3}\\
\{\theta^{a}_{I}(\mathbf{x}),\sigma_{J}^{b}(\mathbf{y})\}&=&m^{2}\varepsilon^{0ab}\eta_{IJ}\delta^{2} (\mathbf{x}-\mathbf{y}),\label{pri4}
\end{eqnarray}
and also the non-vanishing Poisson brackets between primary and secondary constraints presented in Appendix \ref{Algebra0}.  On using (\ref{ConsistencyPrimary}), (\ref{pri1})-(\ref{pri4}) and (\ref{Z1})-(\ref{Z2}),  we find that the consistency condition for each primary constraint becomes
\begin{eqnarray}
\dot{\phi}^{a}_{I}&=&\varepsilon^{0ab}\left(-\frac{1}{m^{4}}\epsilon_{IJK}f^{J}_{0}f^{K}_{b}+\frac{1}{m^{2}}\epsilon_{IJK}w^{J}_{0}f^{K}_{b}-\nu\epsilon_{IJK}e^{J}_{0}e^{K}_{b}+\mathrm{D}_{b}w_{0I}-\frac{1}{m^{2}}\mathrm{D}_{b}f_{0I}+\frac{1}{m^{2}}\lambda_{bI}-\upsilon_{bI}\right)\approx0,\\
\dot{\sigma}^{a}_{I}&=&m^{2}\varepsilon^{0ab}\left(-\mathrm{D}_{b}h_{0I}-\frac{\nu}{m^{2}}\epsilon_{IJK}f^{J}_{0}e^{K}_{b}+\epsilon_{IJK}w^{J}_{0}h^{K}_{b}-\frac{\nu}{m^{2}}\epsilon_{IJK}e^{J}_{0}f^{K}_{b}-\frac{2}{\mu}\nu m^{2}\epsilon_{IJK}e^{J}_{0}e^{K}_{b}+\iota_{bI}\right)\approx0,\\
\dot{\psi}^{a}_{I}&=&\varepsilon^{0ab}\left(m^{2}\epsilon_{IJK}h^{J}_{0}e^{K}_{b}+m^{2}\epsilon_{IJK}e^{J}_{0}h^{K}_{b}+\mathrm{D}_{b}f_{0I}+\frac{1}{m^{2}}\epsilon_{IJK}f^{J}_{0}f^{K}_{b}-\epsilon_{IJK}w^{J}_{0}f^{K}_{b}+\left(\nu-m^{2}\right)\mathrm{D}_{b}w_{0I}\right.\nonumber\\
&&\left.-\lambda_{bI}-\left(\nu-m^{2}\right)\upsilon_{bI}\right)\approx0,\\
\dot{\theta}^{a}_{I}&=&m^{2}\varepsilon^{0ab}\left(\epsilon_{IJK}w^{J}_{0}e^{K}_{b}-\mathrm{D}_{b}e_{0I}+\vartheta_{bI}\right)\approx0.
\end{eqnarray}
These conditions allow us to determine the Lagrange multipliers $\upsilon_{a}^{I}$, $\lambda_{a}^{I}$, $\iota_{a}^{I}$, $ \vartheta_{a}^{I}$, which turn out to be
\begin{eqnarray}
\upsilon_{a}^{I}&\approx&\mathrm{D}_{a}w_{0}^{I}-m^{2}\epsilon^{IJK}e_{0J}e_{aK}+\frac{m^{2}}{\nu}\epsilon^{IJK}h_{0J}e_{aK}+\frac{m^{2}}{\nu}\epsilon^{IJK}e_{0J}h_{aK},\label{multiplier1}\\
\lambda_{a}^{I}&\approx&\mathrm{D}_{a}f_{0}^{I}+m^{2}\left(\nu-m^{2}\right)\epsilon^{IJK}e_{0J}e_{aK}+\frac{m^{4}}{\nu}\epsilon^{IJK}h_{0J}e_{aK}+\frac{m^{4}}{\nu}\epsilon^{IJK}e_{0J}h_{aK}+\frac{1}{m^{2}}\epsilon^{IJK}f_{0J}f_{aK}\nonumber\\
&&-\epsilon^{IJK}w_{0J}f_{aK},\label{multiplier2}\\
\iota_{a}^{I}&\approx&\mathrm{D}_{a}h_{0}^{I}+\frac{\nu}{m^{2}}\epsilon^{IJK}f_{0J}e_{aK}-\epsilon^{IJK}w_{0J}h_{aK}+\frac{\nu}{m^{2}}\epsilon^{IJK}e_{0J}f_{aK}+\frac{2}{\mu}\nu m^{2}\epsilon^{IJK}e_{0J}e_{aK},\label{multiplier3}\\
\vartheta_{a}^{I}&\approx&\mathrm{D}_{a}e_{0}^{I}-\epsilon^{IJK}w_{0J}e_{aK}.\label{multiplier4}
\end{eqnarray}
Substituting back in the primary Hamiltonian (\ref{HamiltonPrimery}) and integrating by parts, we end up with the so-called secondary Hamiltonian:
\begin{equation}
\label{HamiltonianS}
H_{\text{S}}=\int d x^{2}\left(h_{0}^{I}\overline{\Theta}_{I}-f_{0}^{I}\overline{\Phi}_{I}-w_{0}^{I}\overline{\Psi}_{I}-e_{0}^{I}\overline{\Sigma}_{I}\right),
\end{equation}
where $\overline{\Phi}_{I}$, $\overline{\Psi}_{I}$, $\overline{\Sigma}_{I}$, $\overline{\Theta}_{I}$ turn out to be the modificated form of the secondary constraints:
\begin{eqnarray}
\overline{\Phi}_{I}&=&\Phi_{I}+\mathrm{D}_{a}\phi_{I}^{a}+\frac{1}{m^{2}}\epsilon_{IJK}\phi_{a}^{J}f^{aK}+\frac{\nu}{m^{2}}\epsilon_{IJK}\theta_{a}^{J}e^{aK}\approx0,\label{PhiBarra}\\
\overline{\Psi}_{I}&=&\Psi_{I}+\mathrm{D}_{a}\psi^{a}_{I}-\epsilon_{IJK}\phi_{a}^{J}f^{aK}-\epsilon_{IJK}\theta_{a}^{J}h^{aK}-\epsilon_{IJK}\sigma_{a}^{J}e^{aK}\approx0,\label{PsiBarra}\\
\overline{\Sigma}_{I}&=&\Sigma_{I}+\mathrm{D}_{a}\sigma^{a}_{I}+\frac{\nu}{m^{2}}\epsilon_{IJK}\theta_{a}^{J}f^{aK}+2\frac{\nu}{\mu}m^{2}\epsilon_{IJK}\theta_{a}^{J}e^{aK}+\frac{m^{2}}{\nu}\epsilon_{IJK}\psi_{a}^{J}h^{aK}\nonumber\\
&&-m^{2}\epsilon_{IJK}\psi_{a}^{J}e^{aK}+\frac{m^{4}}{\nu}\epsilon_{IJK}\phi_{a}^{J}h^{aK}+m^{2}\left(\nu-m^{2}\right)\epsilon_{IJK}\phi_{a}^{J}e^{aK}\approx0,\label{Sigmabarra}\\
\overline{\Theta}_{I}&=&\Theta_{I}-\mathrm{D}_{a}\theta^{a}_{I}-\frac{m^{4}}{\nu}\epsilon_{IJK}\phi_{a}^{J}e^{aK}-\frac{m^{2}}{\nu}\epsilon_{IJK}\psi_{a}^{J}e^{aK}\approx0,\label{ThetaBarra}
\end{eqnarray}
and $h_{0}^{I}$, $f_{0}^{I}$, $w_{0}^{I}$, $e_{0}^{I}$ remain Lagrange multipliers.  Before proceeding further, it is interesting to note that, following \cite{banados}, we can replace the constraints $\Phi_{I},\Psi_{I},\Sigma_{I},\Theta_{I}$, by the equivalent set of constraints
\begin{equation}
\quad\overline{\Phi}_{I}\approx0,\quad \overline{\Psi}_{I}\approx0,\quad\overline{\Sigma}_{I}\approx0,\quad\overline{\Theta}_{I}\approx0.\label{SecondMod}\nonumber
\end{equation}
It is physically acceptable due to the fact that the constraint surface defined by $\Phi_{I}$, $\Psi_{I}$, $\Sigma_{I}$, $\Theta_{I}$, $\phi^{a}_{I}$, $\psi^{a}_{I}$, $\sigma^{a}_{I}$, and $\theta^{a}_{I}$, is equivalent to the surface defined by $\overline{\Phi}_{I}$, $\overline{\Psi}_{I}$, $\overline{\Sigma}_{I}$, $\overline{\Theta}_{I}$, $\phi^{a}_{I}$, $\psi^{a}_{I}$, $\sigma^{a}_{I}$, and $\theta^{a}_{I}$.

In this stage, a sub-manifold $\Gamma_{\text{S}}\subset\Gamma_{P}$ defines the secondary constraint surface, where all constraints discovered until that moment vanish. Analogous to the primary constraints case before, we should ensure that these modified secondary constraints (\ref{PhiBarra})-(\ref{ThetaBarra}) evolve on $\Gamma_{\text{S}}$, that is
 \begin{equation}
\dot{\overline{\Phi}}_{I}=\{\overline{\Phi}_{I},H_{\text{S}}\}\approx0, \quad\dot{\overline{\Psi}}_{I}=\{\overline{\Psi}_{I},H_{\text{S}}\}\approx0, \quad\dot{\overline{\Sigma}}_{I}=\{\overline{\Sigma}_{I},H_{\text{S}}\}\approx0, \quad\dot{\overline{\Theta}}_{I}=\{\overline{\Theta}_{I},H_{\text{S}}\}\approx0, 
\end{equation}
where $H_{\text{S}}$ is given in Eq (\ref{HamiltonianS}). To study stability conditions of the modified secondary constraints, we need to calculate the Poisson brackets of all the secondary constraints among themselves and with the primary ones. In Appendix \ref{Algebra1}, we show that all the Poisson brackets between primary and modified secondary constraints vanished weakly. While the only non-vanishing Poisson brackets among all the secondary constraints are given by (see appendix \ref{Algebra2} for details)
\begin{eqnarray}
\{\overline{\Phi}_{I}(\mathbf{x}),\overline{\Phi}_{J}(\mathbf{y})\}&\approx& \frac{\nu}{m^{2}}\varepsilon^{0ab}e_{aI}e_{bJ}\delta^{2}(\mathbf{x}-\mathbf{y}),\\
\{\overline{\Phi}_{I}(\mathbf{x}),\overline{\Sigma}_{J}(\mathbf{y})\}&\approx&\frac{\nu}{m^{2}}\epsilon_{I}{^{KN}}\epsilon_{JK}{^{M}}\varepsilon^{0ab}f_{aN}e_{bM}\delta^{2}(\mathbf{x}-\mathbf{y}),\\
\{\overline\Theta_{I}(\mathbf{x}),\overline\Theta_{J}(\mathbf{y})\}&\approx&-\frac{m^{4}}{\nu}\varepsilon^{0ab}e_{aI}e_{bJ}\delta^{2}(\mathbf{x}-\mathbf{y}),\\
\{\overline\Theta_{I}(\mathbf{x}),\overline\Sigma_{J}(\mathbf{y})\}&\approx&\frac{m^{4}}{\nu}\epsilon_{I}{^{KN}}\epsilon_{JK}{^{M}}\varepsilon^{0ab}h_{aN}e_{bM}\delta^{2}(\mathbf{x}-\mathbf{y}),\\
\{\overline\Sigma_{I}(\mathbf{x}),\overline\Sigma_{J}(\mathbf{y})\}&\approx&\left(-\frac{\nu}{m^{2}}\varepsilon^{0ab}f_{aI}f_{bJ}-\frac{m^{4}}{\nu}\varepsilon^{0ab}h_{aI}h_{bJ}\right)\delta^{2}(\mathbf{x}-\mathbf{y}).
\end{eqnarray}
Using this information, we find that the stability under time evolution of the modified secondary constraints implies the following set of integrability conditions:
\begin{eqnarray}
\dot{\overline{\Phi}}_{I}&\approx&\frac{\nu}{m^{2}}\varepsilon^{\alpha\beta\mu}e_{\alpha I}e_{\beta J}f^{J}_{\mu}\approx0,\\
\dot{\overline{\Theta}}_{I}&\approx&\frac{m^{4}}{\nu}\varepsilon^{\alpha\beta\mu}e_{\alpha I}h_{\beta J}e^{J}_{\mu}\approx0,\\
\dot{\overline{\Sigma}}_{I}&\approx&\frac{m^{4}}{\nu}\varepsilon^{\alpha\beta\mu}h_{\alpha I}h_{\beta J}e^{J}_{\mu}+\frac{\nu}{m^{2}}\varepsilon^{\alpha\beta\mu}f_{\alpha I}f_{\beta J}e^{J}_{\mu}\approx0.
\end{eqnarray}
If we only demand  that the dreibein $e_{\alpha}^{I}$ be invertible (a requirement for first-order theory to be equivalent to the standard second-order formulation of gravity) then we have $6$ constraint aquations  in the following form:
\begin{eqnarray}
\Xi^{\alpha}&=&\varepsilon^{\alpha\beta\mu}e_{\beta I}f^{I}_{\mu}\approx0,\label{terc1}\\
\Upsilon^{\alpha}&=&\varepsilon^{\alpha\beta\mu}e_{\beta I}h^{I}_{\mu}\approx0.\label{terc2}
\end{eqnarray}
 These conditions, known as the symmetrization conditions, are crucial to show the equivalence between the metric and  first-order formulation of massive gravity theories \cite{Deffayet}. It is straightforward to see that the above equations Eqs. (\ref{terc1})-(\ref{terc2}) can be decomposed  in
\begin{eqnarray}
\Xi&=&\varepsilon^{0ab}e_{a I}f^{I}_{b}\approx0,\label{ter1}\\
\Xi^{a}&=&\varepsilon^{0ab}\left(e_{bJ}f^{J}_{0}-e_{0J}f^{J}_{b}\right)\approx0,\label{ter2}\\
\Upsilon&=&\varepsilon^{0ab}e_{a I}h^{I}_{b}\approx0,\label{ter3}\\
\Upsilon^{a}&=&\varepsilon^{0ab}\left(e_{bJ}h^{J}_{0}-e_{0J}h^{J}_{b}\right)\approx0. \label{ter4}
\end{eqnarray} 
One straightforwardly verifies that the functions (\ref{ter2}) and (\ref{ter4}) mix dynamical variables $(e_{a}^{I} , f_{a}^{I} , h_{a}^{I})$ with Lagrangian multipliers $(e_{0}^{I} , f_{0}^{I} , h_{0}^{I})$, consequently these relations are not constraints. Whereas the functions (\ref{ter1}) and (\ref{ter3}) constitute a new set of two tertiary constraints of the theory, as they are algebraic relations involving only the canonical variables (not the Lagrange multipliers). Clearly, all the constraints derived until this point define the tertiary constraint surface $\Gamma_{\text{T}}\subset\Gamma_{S}$. Now,  when introducing these tertiary constraints Eqs. (\ref{ter1}) and (\ref{ter3}) into the secondary Hamiltonian (\ref{HamiltonianS}) through the Lagrange multipliers $u$ and $z$, we arrive at the following tertiary Hamiltonian:
\begin{equation}
H_{\text{T}}=\int dx^{2}\left(h_{0}^{I}\overline{\Theta}_{I}-f_{0}^{I}\overline{\Phi}_{I}-w_{0}^{I}\overline{\Psi}_{I}-e_{0}^{I}\overline{\Sigma}_{I}+u\Xi+z\Upsilon\right).\label{Hamilton3}
\end{equation}
The next step in the Dirac algorithm is to demand the preservation in time of each tertiary constraint (\ref{ter1}) and (\ref{ter3}). First, it should be obvious that all the Poisson brackets among tertiary constraints vanish identically because neither of the functionals depend on the momenta. Second, the non-trivial Poisson brackets between primary and tertiary constraints are given by
\begin{eqnarray}
\{\phi^{a}_{I}(\mathbf{x}),\Xi(\mathbf{y})\}&=&\varepsilon^{0ab}e_{bI}\delta^{2}(\mathbf{x}-\mathbf{y}),\\
\{\sigma^{a}_{I}(\mathbf{x}),\Xi(\mathbf{y})\}&=&-\varepsilon^{0ab}f_{bI}\delta^{2}(\mathbf{x}-\mathbf{y}),\\
\{\sigma^{a}_{I}(\mathbf{x}),\Upsilon(\mathbf{y})\}&=&-\varepsilon^{0ab}h_{bI}\delta^{2}(\mathbf{x}-\mathbf{y}),\\
\{\theta^{a}_{I}(\mathbf{x}),\Upsilon(\mathbf{y})\}&=&\varepsilon^{0ab}e_{bI}\delta^{2}(\mathbf{x}-\mathbf{y}).
\end{eqnarray}
Finally, the non-vanishing Poisson brackets between secondary and tertiary constraints are found as (see Appendix \ref{Algebra3} for details)
\begin{eqnarray}
\{\overline\Phi_{I}(\mathbf{x}),\Xi(\mathbf{y})\}&\approx&-\frac{1}{m^{2}}\varepsilon^{0ab}\epsilon_{IJK}e_{a}^{J}f_{b}^{K}\delta^{2}(\mathbf{x}-\mathbf{y}),\\
\{\overline\Theta_{I}(\mathbf{x}),\Xi(\mathbf{y})\}&\approx&\frac{m^{4}}{\nu}\varepsilon^{0ab}\epsilon_{IJK}e^{J}_{a}e^{K}_{b}\delta^{2}(\mathbf{x}-\mathbf{y}),\\
\{\overline\Sigma_{I}(\mathbf{x}),\Xi(\mathbf{y})\}&\approx&-\varepsilon^{0ab}\epsilon_{IJK}\left(\frac{3}{2}m^{2}\left(\nu-m^{2}\right)e_{a}^{J}e_{b}^{K}+ 2\frac{m^{4}}{\nu}h_{a}^{J}e_{b}^{K}+\frac{1}{2m^{2}}f_{a}^{J}f_{b}^{K}\right)\delta^{2}(\mathbf{x}-\mathbf{y}),\\
\{\overline\Phi_{I}(\mathbf{x}),\Upsilon(\mathbf{y})\}&\approx&-\frac{\nu}{m^{2}}\varepsilon^{0ab}\epsilon_{IJK}e^{J}_{a}e^{K}_{b}\delta^{2}(\mathbf{x}-\mathbf{y}),\\
\{\overline\Sigma_{I}(\mathbf{x}),\Upsilon(\mathbf{y})\}&\approx&-\varepsilon^{0ab}\epsilon_{IJK}\left(2\frac{\nu}{m^{2}}f_{a}^{J}e_{b}^{K}+3\frac{\nu}{\mu}m^{2}e_{a}^{J}e_{b}^{K}\right)\delta^{2}(\mathbf{x}-\mathbf{y}).
\end{eqnarray}
Thus, the consistency condition on  the  tertiary constraints  yields two equations of the following form:
\begin{eqnarray}
\dot{\Upsilon}&\approx&\frac{\nu}{m^{2}}\epsilon^{\alpha\beta\mu}\epsilon_{IJK}e_{\alpha}^{I}f^{J}_{\beta}e_{\mu}^{K}+\frac{\nu m^{2}}{\mu}\epsilon^{\alpha\beta\mu}\epsilon_{IJK}e_{\alpha}^{I}e^{J}_{\beta}e_{\mu}^{K}\approx0,\label{quartic1}\\
\dot{\Xi}&\approx&-\frac{m^{4}}{\nu}\varepsilon^{\alpha\beta\mu}\epsilon_{IJK}e_{\alpha}^{I}h_{\beta}^{J}e_{\mu}^{K}-\frac{1}{2m^{2}}\varepsilon^{\alpha\beta\mu}\epsilon_{IJK}f_{\alpha}^{I}e_{\beta}^{J}f_{\mu}^{K}-\frac{1}{2}m^{2}\left(\nu-m^{2}\right)\varepsilon^{\alpha\beta\mu}\epsilon_{IJK}e_{\alpha}^{I}e_{\beta}^{J}e_{\mu}^{K}\approx0.\label{quartic2}
\end{eqnarray}
Notice that, these expressions do not define new constraints, because both mix dynamical variables with Lagrange multipliers. However, such conditions present themselves in two scalar equations, establishing the most general relationship between the fields defining the EGMG theory in the first-order formulation. Hence the preservation of the tertiary constraints (\ref{ter1}) and (\ref{ter3}) do not give rise to any quartic constraint. Such that the number of constraints in the system is now closed and our method to obtain the true set of constraints has terminated. Consequently, we have the complete set of constraints on the dynamics of the theory.

We now focus our attention on (\ref{quartic1}) and (\ref{quartic2}). Let us  define the following objects 
\begin{equation}
\Lambda_{\alpha}{^{\beta}}=e_{\alpha}^{I}f^{\beta}_{I}\quad\&\quad\Delta_{\alpha}{^{\beta}}=e^{I}_{\alpha}h^{\beta}_{I},
\end{equation} 
and consider the identity $\epsilon^{\alpha\beta\mu}e_{\alpha}^{I}e^{J}_{\beta}e_{\mu}^{K}={\bf e}\epsilon^{IJK}$ with ${\bf e}=\text{det}|e_{\alpha}^{I}|\neq0$. In doing so, and after some algebra, we find that the Eqs. (\ref{quartic1}) and (\ref{quartic2}) amount to
\begin{eqnarray}
\dot{\Upsilon}&\approx&-2{\bf e}\frac{\nu}{m^{2}}\left(\Lambda+3\frac{m^{4}}{\mu}\right)\approx0,\label{AA}\\
\dot{\Xi}&\approx&2{\bf e}\frac{m^{4}}{\nu}\left(\Delta+\frac{\nu}{4m^{6}}\left(\Lambda^{2}-\Lambda_{\alpha}^{\beta}\Lambda_{\beta}^{\alpha}\right)+\frac{3\nu}{2m^{2}}\left(\nu-m^{2}\right)\right)\approx0,\label{B}
\end{eqnarray}
where $\Lambda=\Lambda_{\alpha}{^{\alpha}}$ and $\Delta=\Delta_{\alpha}{^{\alpha}}$.  It is now straightforward to see that the right-hand side of Eq. (\ref{AA}) will be  equal to zero on the tertiary constraint surface only if the following condition holds true in $\Gamma_{T}$: 
\begin{equation}
\Lambda\approx-3\frac{m^{4}}{\mu}.
\end{equation}
Therefore the auxiliary field $f^{\alpha}_{I}$ turns out to be proportional to the dreibein $e^{\alpha}_{I}$, i.e.
\begin{equation}
f^{\alpha}_{I}\approx c_{f}e^{\alpha}_{I},\quad\text{with}\quad c_{f}=-\frac{m^{4}}{\mu}.\label{F}
\end{equation}
Substituting Eq. (\ref{F}) in Eq. (\ref{B}),  we find that
\begin{equation}
\Delta\approx-3\frac{\nu}{2}\left(\frac{m^{2}}{\mu^{2}}+\frac{\left(\nu-m^{2}\right)}{m^{2}}\right),
\end{equation}
It immediately implies that the auxiliary field $h^{\alpha}_{I}$ is also proportional to the dreibein $e^{\alpha}_{I}$ in $\Gamma_{T}$. The final solution is
\begin{equation}
h^{\alpha}_{I}\approx c_{h}e^{\alpha}_{I},\quad\text{with}\quad c_{h}=-\frac{\nu}{2}\left(\frac{m^{2}}{\mu^{2}}+\frac{\left(\nu-m^{2}\right)}{m^{2}}\right).\label{FyH}
\end{equation}
 It is worthwhile noticing that the consistency under time evolution of the tertiary constraints allows us to fix the value of the auxiliary fields, $(f^{I}_{\mu},h^{I}_{\mu})\approx(c_{f},c_{h})e^{I}_{\mu}$, but only in the tertiary constraint surface $\Gamma_{T}$ (and not in the entire phase-space).  In this case, we can decompose such fields into its space and time components,
\begin{eqnarray}
f^{I}_{\mu}&\longrightarrow&\left(f^{I}_{0},f^{I}_{a}\right)\approx c_{f}(e^{I}_{0},e^{I}_{a}),\label{propor}\\
h^{I}_{\mu}&\longrightarrow&\left(h^{I}_{0},h^{I}_{a}\right)\approx c_{h}(e^{I}_{0},e^{I}_{a}).\label{propor2}
\end{eqnarray}
With all this information, it is remarkable that if the temporal component, $f^{0}_{I}$ and $h^{0}_{I}$, of the fields $f^{\alpha}_{I}$ and $h^{\alpha}_{I}$ is inserted into the tertiary Hamiltonian  (\ref{Hamilton3}),  then a new Hamiltonian is defined,
\begin{equation}
H_{\text{N}}=\int d x^{2}\left(-e_{0}^{I}\widehat{\Sigma}_{I}-w_{0}^{I}\overline{\Psi}_{I}+u\Xi+z\Upsilon\right),\label{prefinal}
\end{equation}
where $\widehat{\Sigma}_{I}$ is now defined by
\begin{equation}
\widehat{\Sigma}_{I}=\overline{\Sigma}_{I}+c_{f}\overline{\Phi}_{I}-c_{h}\overline{\Theta},
\end{equation}
whereas $\overline{\Psi}_{I}$ remains as in Eq. (\ref{PsiBarra}). 

Now, below we summarize  the whole set of constraints obtained in the above  analysis for reference:
\begin{eqnarray}
\text{24\quad Primary constraints}&:& \phi^{a}_{I}, \quad\psi^{a}_{I}, \quad\sigma^{a}_{I}, \quad \theta^{a}_{I}.\label{FPrimary}\\
\text{12\quad Secondary constraints}&:& \overline{\Phi}_{I}, \quad\overline{\Psi}_{I},\quad \widehat{\Sigma}_{I}, \quad\overline{\Theta}_{I}.\label{Fsecondary}\\
\text{2\quad Tertiary constraints}&:&\Xi,\quad\Upsilon.\label{Ftertiary}
\end{eqnarray}
To conclude this section, it is worth remembering that all these constraints define the tertiary constraint surface. Besides, as we previously said, consistency of the Hamiltonian analysis requires us to not only enforce the whole set of constraints but also to guarantee that each constraint appearing in the theory has to be preserved during the dynamical evolution of the system. Thus, this requirement implies that the Lagrange multipliers $u$ and $z$ must be readily zero. In this way, we can say that the consistency of all the constraints under time evolution is dynamically ensured, and so the Dirac algorithm closes here. Consequently, we find the following form of the final Hamiltonian:
 \begin{equation}
 H_{\text{F}}=-\int d x^{2}\left(e_{0}^{I}\widehat{\Sigma}_{I}+w_{0}^{I}\overline{\Psi}_{I}\right),\label{HamiltonianFirstClasss}
 \end{equation}
which shows that the system is totally constrained so that the constraints generate the time evolution. In this configuration, we will see that $\widehat{\Sigma}_{I}$and $\overline{\Psi}_{I}$ are the $6$ first-class constraints corresponding to the six Lagrange multipliers that remain unfixed.

\section{First- and second-class constraints}
\label{classify}
\subsection{ Separation of constraints and physical degrees of freedom}
Having determined all the constraints in the theory, the next step in the Hamiltonian analysis is the classification of all the constraints into first- and second-class, which requires computing the various Poisson brackets between all the constraints.  In general, the so-called first-class constraints are characterized by the property that only they weakly commute with all the constraints of the system.  On the other hand, the constraints that have at least one weakly nonvanishing Poisson bracket are referred to as second-class. As a consequence, according to Dirac's conjecture, each first-class constraint generates a gauge symmetry on the constraint surface. Furthermore, the second-class constraints eliminate non-physical fields not related to the symmetries and also allow us to build  a new symplectic structure, the Dirac's brackets, to quantize a gauge system. 

With the help of the constraints algebra reported in Appendix \ref{Algebra31}, we see that all Poisson brackets of the constraints $\widehat{\Sigma}_{I}$ and $\overline{\Psi}_{I}$ vanish in the tertiary constraint surface and hence they prove to be first-class constraints, that is
\begin{equation}
\text{6 \quad First-class constraints : } \widehat{\Sigma}_{I}\approx0,\quad\overline{\Psi}_{I}\approx0.
\end{equation}
As a consequence, the final Hamiltonian in Eq. (\ref{HamiltonianFirstClasss}) must be first-class, and therefore it has the effect of combining dynamics with gauge transformations. 

On the other hand, it turns out that the Poisson brackets among the remaining constraints do not vanish, indicating the fact that the theory has the following set of second-class constraints:
\begin{equation}
\text{32 \quad Second-class constraints : } \phi^{a}_{I}\approx0, \quad\psi^{a}_{I}\approx0, \quad\sigma^{a}_{I}\approx0, \quad \theta^{a}_{I}\approx0,\quad\overline{\Phi}_{I}\approx0,\quad\overline{\Theta}_{I}\approx0,\quad\Xi\approx0,\quad\Upsilon\approx0.
\end{equation}

We are now in a position to count the number of physical degrees of freedom of the given theory from the following formula \cite{Henneaux},
\begin{equation}
\mathcal{N}=\frac{1}{2}\left(\mathcal{P}-2\mathcal{F}-\mathcal{S}\right),
\end{equation}
where $\mathcal{P}$ is number of phase-space variables $\left(f_{a}^{I},\Pi_{J}^{b}, w_{a}^{I},\pi_{J}^{b}, e_{a}^{I},\widetilde{\Pi}_{J}^{b}, h_{a}^{I},\widetilde{\pi}_{J}^{b}\right)$, $\mathcal{F}$ is the number of first-class constraints $\left(\widehat{\Sigma}_{I},\overline{\Psi}_{I}\right)$, and  $\mathcal{S}$ is the number of second-class constraints $\left(\phi^{a}_{I}, \psi^{a}_{I}, \sigma^{a}_{I}, \theta^{a}_{I},\overline{\Phi}_{I},\overline{\Theta}_{I},\Xi,\Upsilon\right)$ \cite{Henneaux}. Hence, it is concluded that  exotic general massive gravity  possesses $1/2 (48-2\times 6-32)=2$ physical degrees of freedom per space point corresponding to one massive graviton in three dimensions. In other words, we have shown that the physical phase-space of EGMG in the first-order formulation has dimension $4$ per space point and that it is ghost free which is the desired result in agreement with \cite{EMG}. It is worth commenting that, the number of local DoF was computed also in the original article of EGMG \cite{EMG}, by using a minimal Hamiltonian formalism. Although the number of DoF obtained in [32] is correct, some aspects of the corresponding derivations are not satisfying: they are based on introducing two ad-hoc constraints by appealing to the Lagrangian equations, but the effect of this procedure on the overall constraint structure of the theory remains unclear.

\subsection{Dirac brackets}
\label{Dirac barcket}
Once the whole set of second-class constraints $\chi^{A}=\left(\phi^{a}_{I},\psi^{a}_{I},\sigma^{a}_{I}, \theta^{a}_{I},\overline{\Phi}_{I},\overline{\Theta}_{I},\Xi,\Upsilon\right)$ has been identified, it can be eliminated from the theory by defining  a new symplectic structure for the system, which is  called the Dirac bracket.  In order to see this, let us define the Dirac matrix  $ \boxplus^{AB}$ whose elements are the Poisson brackets among these second-class constraints, i.e.
\begin{equation}
 \boxplus^{AB}(\mathbf{x},\mathbf{y})=\{\chi^{A}(\mathbf{x}),\chi^{B}(\mathbf{y})\},\quad \text{with}\quad \boxplus^{AB}=-\boxplus^{BA}.
\end{equation}
Now let's observe that the Dirac matrix can be written in block matrix form as
\begin{equation}
 \boxplus^{AB}(\mathbf{x},\mathbf{y})=
\label{Ff}\begin{pmatrix}
 \mathbb{A}  &  \mathbb{B}  \\                                                                        
 \mathbb{C}   &  \mathbb{D} 
\end{pmatrix}\delta^{2}(\mathbf{x}-\mathbf{y}),\quad\text{with}\quad\mathbb{C}=-\mathbb{B} ^{T}.
\end{equation}
In this case, the explicit form of each sub-matrix $\mathbb{A}$, $\mathbb{B}$ and $\mathbb{D}$ in Eq. (\ref{Ff}) turns out to be
\begin{eqnarray}
\label{A}
\mathbb{A}&=&\varepsilon^{0ab}
\begin{pmatrix}
 1/m^{2}      &  -1   &  0    &  0    \\                                                                        
1    &  -\left(\nu-m^{2}\right)      &  0   &   0   \\
    0     &  0    & 0  & m^{2} \\
0  &   -m^{2}  &0  & 0 	
\end{pmatrix}\eta^{IJ},
\end{eqnarray}
\begin{eqnarray}
\mathbb{B}&=&\varepsilon^{0ab}
\begin{pmatrix}
 0      &  0   &  e_{bI}    &  0    \\                                                                        
0    & 0      &  0   &   0   \\
    0     &  0    & -f_{bI}  & -h_{bI} \\
0  &   0  &0  & e_{bI} 	
\end{pmatrix},
\end{eqnarray}
\begin{eqnarray}
\label{D}
\mathbb{D}=\varepsilon^{0ab}
\begin{pmatrix}
 (\nu/m^{2})e_{aI}e_{bJ}     &  0   &  -(1/m^{2})\epsilon_{IJK}e_{a}^{J}f_{b}^{K}    &  -(\nu/m^{2})\epsilon_{IJK}e_{a}^{J}e_{b}^{K}    \\                                                                        
0    & -(m^{4}/\nu)e_{aI}e_{bJ}      &  (m^{4}/\nu)\epsilon_{IJK}e_{a}^{J}e_{b}^{K}   &   0   \\
    (1/m^{2})\epsilon_{IJK}e_{a}^{J}f_{b}^{K}     &  -(m^{4}/\nu)\epsilon_{IJK}e_{a}^{J}e_{b}^{K}    & 0  & 0 \\
(\nu/m^{2})\epsilon_{IJK}e_{a}^{J}e_{b}^{K}  &   0  &0  & 0
\end{pmatrix}.
\end{eqnarray}
From (\ref{A}), it is easy to see that $\mathbb{A}$ is invertible. Since the Dirac matrix is of block form, it can be inverted blockwise on the tertiary constraint surface as follows (see appendix \ref{DB} for details):
\begin{equation}
\label{symplectic}
\left( \boxplus^{AB}\right)^{-1}=
\begin{pmatrix}
 \mathbb{A}^{-1}+\mathbb{A}^{-1}\mathbb{B}\left(\mathbb{D}-\mathbb{C}\mathbb{A}^{-1}\mathbb{B}\right)^{-1}\mathbb{C}\mathbb{A}^{-1} & \quad -\mathbb{A}^{-1}\mathbb{B}\left(\mathbb{D}-\mathbb{C}\mathbb{A}^{-1}\mathbb{B}\right)^{-1}  \\                                                                        
-\left(\mathbb{D}-\mathbb{C}\mathbb{A}^{-1}\mathbb{B}\right)^{-1}\mathbb{C}\mathbb{A}^{-1}   &\quad \left(\mathbb{D}-\mathbb{C}\mathbb{A}^{-1}\mathbb{B}\right)^{-1}
\end{pmatrix}\delta^{2}(\mathbf{x}-\mathbf{y}).
\end{equation}

For any two functions of canonical variables, say $\mathcal{O}_{1}$ and $\mathcal{O}_{A}$, the Dirac brackets for this system are defined by \cite{P. A. M}
\begin{equation}
\{\mathcal{O}_{1}(\mathbf{x}),\mathcal{O}_{2}(\mathbf{y})\}_{D}=\{\mathcal{O}_{1}(\mathbf{x}),\mathcal{O}_{2}(\mathbf{y})\}-\int dx^{2}dy^{2}\{\mathcal{O}_{1}(\mathbf{x}),\mathcal{S}_{A}(\mathbf{z})\}\left( \boxplus^{AB}(\mathbf{z},\mathbf{w})\right)^{-1}\{\mathcal{S}_{B}(\mathbf{w}),\mathcal{O}_{2}(\mathbf{y})\},\label{FormDB}
\end{equation}
where $\{\mathcal{O}_{1}(\mathbf{x}),\mathcal{O}_{2}(\mathbf{y})\}$ is the Poisson bracket between $\mathcal{O}_{1}$ and $\mathcal{O}_{A}$. Using this information, we proceed to compute the Dirac's brackets among all the variables of the full phase space $\Gamma$ in Appendix \ref{DB}. After constructing the corresponding Dirac brackets, the second-class constraints can be used as strong equalities, i.e. as identities expressing some variables in terms of others. As a result, we discover that the Dirac's brackets between the dynamical variables parametrizing the physical phase space are given by
\begin{eqnarray}
\{e_{I}^{a}(\mathbf{x}),e^{b}_{J}(\mathbf{y})\}_{\text{D}}&=&\frac{1}{2}\frac{m^{2}}{c_{f}^{2}}\frac{m^{2}-\nu}{2m^{2}-\nu}\varepsilon^{0ab}\eta_{IJ}\delta^{2}(\mathbf{x}-\mathbf{y}),\label{DB1}\\
\{e_{I}^{a}(\mathbf{x}),w^{b}_{J}(\mathbf{y})\}_{\text{D}}&=&\frac{1}{2}\frac{m^{2}}{c_{f}}\frac{1}{\left(2m^{2}-\nu\right)}\varepsilon^{0ab}\eta_{IJ}\delta^{2}(\mathbf{x}-\mathbf{y}),\label{DB2}\\
\{w_{I}^{a}(\mathbf{x}),w^{b}_{J}(\mathbf{y})\}_{\text{D}}&=&\frac{1}{\left(2m^{2}-\nu\right)}\varepsilon^{0ab}\eta_{IJ}\delta^{2}(\mathbf{x}-\mathbf{y}).\label{DB3}
\end{eqnarray}
As we see, these Dirac brackets do not have a common canonical structura, in fact, the structure becomes non-commutative. In addition, the results of these Dirac brackets are different from those of standard Poisson brackets in that the information on the constrained dynamics of the EGMG is apparent. These Dirac brackets could be useful to study physical observables, as well as, for performing the quantization of the theory. According to Dirac's prescription \cite{Dirac,Henneaux,P. A. M,Blagojevic}, the quantization is carried out via the replacement of  the Dirac brackets with commutators with a factor of $1/i\hbar$,
\begin{equation}
  \{\mathcal{O}_{1}(\mathbf{x}),\mathcal{O}_{2}(\mathbf{y})\}_{D}\longrightarrow\frac{1}{i\hbar}\left[\widehat{\mathcal{O}}_{1}(\mathbf{x}),\widehat{\mathcal{O}}_{2}(\mathbf{y})\right].
  \end{equation}
Furthermore, a physical state $\psi$ must satisfy 
\begin{equation}
\mathbf{\widehat{\mathcal{O}}}|\psi\rangle,
\end{equation}
 where $\widehat{\mathcal{O}}$ is a quantized version of  the first-class constraints or observables.
\subsection{Gauge transformations}
To conclude, we can derive the generator and the local symmetries for EGMG. As we know, gauge invariance is one of the most significant and practical concepts in theoretical physics. The existence of gauge symmetries in the mathematical structure of a given physical theory is the sign of the presence of interactions, in addition to restricting the nature of observable quantities \cite{Regularization}. According to Dirac's conjecture, the most general expression for the generator of correct gauge transformations of the system should be constructed as a linear combination of all first-class constraints of the theory \cite{Dirac}. Thus, the generator of gauge transformations is proposed as,
\begin{equation}
G=\int dx^{2}\left(\alpha^{I}\widehat{\Sigma}_{I}+\beta^{I}\overline{\Psi}_{I}\right),\label{generator}
\end{equation}
where  $\alpha^{I}$ and $\beta^{I}$ are the gauge parameters. Hence, to obtain gauge variation $\delta_{G}$ of every physical variable $\mathcal{A}$ generated by $G$, we can use the Poisson bracket  of the corresponding variable with the generating functional via the following relation:
\begin{equation}
\delta_{\text{G}}\mathcal{A}=\{\mathcal{A},G\}.
\end{equation}
Therefore,  this gives rise to the following gauge transformation:
\begin{eqnarray}
\delta_{G}e^{I}_{a}&=&D_{a}\alpha^{I}+\epsilon^{IJK}\beta_{J}e_{aK},\label{G1}\\
\delta_{G}w^{I}_{a}&=&D_{a}\beta^{I}+m^{2}\epsilon^{IJK}\alpha_{J}e_{aK}-\frac{m^{2}}{\nu}\epsilon^{IJK}\alpha_{J}\left(h_{aK}+c_{h}e_{aK}\right)\label{G2},\\
\delta_{G}f^{I}_{a}&=&c_{f}D_{a}\alpha^{I}+\epsilon^{IJK}\beta_{J}f_{aK}+\frac{m^{2}}{\mu}\epsilon^{IJK}\alpha_{J}f_{aK}-m^{2}\left(\nu-m^{2}\right)\epsilon^{IJK}\alpha_{J}e_{aK}-\frac{m^{4}}{\nu}\epsilon^{IJK}\alpha_{J}\left(h_{aK}+c_{h}e_{aK}\right)\label{G3},\\
\delta_{G}h^{I}_{a}&=&c_{h}D_{a}\alpha^{I}+\epsilon^{IJK}\beta_{J}h_{aK}-2\frac{\nu}{\mu}m^{2}\epsilon^{IJK}\alpha_{J}e_{aK}-\frac{\nu}{m^{2}}\epsilon^{IJK}\alpha^{J}\left(f_{aK}+c_{f}e_{aK}\right).\label{G4}
\end{eqnarray}
But these are no diffeomorphisms (diff). Nevertheless,  we can redefine the gauge parameters in terms of the diffeomorphism parameters, where the relations depend on the dynamical variables,
\begin{equation}
\alpha_{I}=\xi_{a}e^{a}_{I},\quad \beta_{I}=-\xi_{a}w^{a}_{I}, \label{parametrization}
\end{equation}
with $\xi_{a}$ an arbitrary three-vector. In this manner, substituting (\ref{parametrization}) in the gauge transformations Eqs. (\ref{G1})-(\ref{G4}) and using the expressions for the spatial component of auxiliary fields given in Eqs.  (\ref{propor})-(\ref{propor2}), we finally obtain the spatial diffeomorphism,  modulo the constraints, for the dynamical variables as it should be, namely,
\begin{eqnarray}
\delta_{\text{diff}} e_{a}^{I}&=&\mathfrak{L}_{\xi}e_{a}^{I}-\varepsilon_{0ab}\xi^{b}\Theta^{I},\\
\delta_{\text{diff}} w_{a}^{I}&=&\mathfrak{L}_{\xi}w_{a}^{I}-\frac{1}{\nu}\varepsilon_{0ab}\xi^{b}\left(\Psi^{I}+m^{2}\Phi^{I}\right),\\
\delta_{\text{diff}} f_{a}^{I}&=&\mathfrak{L}_{\xi}f_{a}^{I}+\frac{m^{2}}{\nu}\varepsilon_{0ab}\xi^{b}\left(\Psi^{I}-\left(\nu-m^{2}\right)\Phi^{I}\right),\\
\delta_{\text{diff}} h_{a}^{I}&=& \mathfrak{L}_{\xi}h_{a}^{I}+\frac{1}{m^{2}}\varepsilon_{0ab}\xi^{b}\Sigma^{I},
\end{eqnarray}
where $\mathfrak{L}_{\xi}$ is a Lie derivative along $\xi$.

\section{ SUMMARY AND DISCUSSIONS}
\label{conclusions}
In this work, we have studied the Hamiltonian analysis of the Exotic General Massive Gravity theory written in the first-order formulation. To obtain the best Hamiltonian description of this model, all the steps of Dirac's framework were followed. Our basic goal was to obtain and classify all the constraints on the dynamics of the model and deduce the number of physical degrees of freedom in the theory. First, we wrote the action in a $(2 + 1)-$dimensional form and found the conjugate momenta of each dynamical field in our theory. This enabled us to write the Hamiltonian and determine the primary constraints. Then we analyzed the requirement of the preservation of these constraints and we derived the corresponding modificated secondary constraints. Subsequently, we determined the conditions when these constraints are preserved and we found the symmetrization conditions corresponding to tertiary constraints, Eqs. (\ref{terc1})-(\ref{terc2}). In the process of completing Dirac's consistency procedure, we discovered two scalar equations (\ref{quartic1})-(\ref{quartic2}) establishing the most general relationship between the fields defining the principle action of the model. Such expressions have been solved algebraically for the auxiliary fields in terms of the dreibein, $(f,h)=(c_{f},c_{h})e$, Eqs. (\ref{F}) and (\ref{FyH}). Thereafter, with the complete structure of the constraints and their Poisson brackets, we classify all the constraints into first and second-class ones. The correct classification of such constraints allowed us to show that there are two physical degrees of freedom corresponding to a massive graviton without ghosts. Further, with the help of first-class constraints, we obtained the gauge generators that yield the spatial diffeomorphism symmetry, under which all physical quantities must be invariant. Finally, we were able to construct the Dirac brackets of EGMG by using the inverse of the so-called Dirac matrix, of which the entries are the Poisson brackets among the second-class constraints. All of these results have not been reported in the literature as far as we know.

To conclude, it is worth noticing that having a consistent Hamiltonian description, that integrates all information about the constraints, represents a key step toward a proper quantization. Thus, we expect that the results obtained in this paper might have important consequences for the investigation of a possible quantization of EGMG. To our knowledge, the quantization could proceed using canonical quantization methods by constructing first a suitable Hilbert space of quantum states on which the quantum Hamiltonian and constraints operators act, and studying the Dirac observables of the corresponding quantum operators \cite{Dirac,Henneaux,P. A. M,Blagojevic}. In particular, the canonical quantization of the EGMG model could be tackled  using the background-independent techniques developed in the loop quantum gravity program \cite{LGQ}. But to do so first requires defining a discrete phase space in terms of holonomies of the connection and fluxes of the dreibein, such that their Dirac (not Poisson) algebra is the holonomy-flux algebra. Afterward, it would be necessary to express the first-class constraints in terms of these holonomy-flux variables. Subsequently, the quantization of the holonomy-flux variables would lead to an irreducible representation on a Kinematical Hilbert space, spanned by the spin-network states, where the first-class constraints are represented by regularized quantum operators, e.g. see \cite{LGQ2}  for more details. While the strategy seems good —even natural— at first, a priori it is not clear that it would be applied straightforwardly; the main difficulty resides in the non-commutativity of the dreibein and the connection with respect to the Dirac's brackets. This would imply quantizing the fluxes and the holonomies so that they satisfy the quantum analog of Eqs.  (\ref{DB1})-(\ref{DB3}).
\section*{ACKNOWLEDGMENTS}
We acknowledge support from Direcci\'on de Apoyo a Docentes, Investigaci\'on y Posgrado (DADIP) de la Universidad de Sonora (Grant No. USO315008080).
\appendix
\section{Algebra among the constraints}
\label{Algebra}
In this appendix we develop the algebra of  all the constraints. 
\subsection{Poisson brackets between primary and secondary constraints}
\label{Algebra0}
The  non-vanishing Poisson brackets between secondary and primary constraints are:
\begin{eqnarray}
\{\Phi_{I}(\mathbf{x}),\phi_{J}^{a}(\mathbf{y})\}&=&\frac{1}{m^{2}}\varepsilon^{0ab}\left(\frac{1}{m^{2}}\epsilon_{IJK}f_{b}^{K}+\partial^{\mathbf{x}}_{b}\eta_{IJ}-\epsilon_{IJK}w^{K}_{b}\right)\delta^{2} (\mathbf{x}-\mathbf{y}),\label{Z1}\\
\{\Psi_{I}(\mathbf{x}),\phi_{J}^{a}(\mathbf{y})\}&=&-\varepsilon^{0ab}\left(\frac{1}{m^{2}}\epsilon_{IJK}f_{b}^{K}+\partial^{\mathbf{x}}_{b}\eta_{IJ}-\epsilon_{IJK}w^{K}_{b}\right)\delta^{2} (\mathbf{x}-\mathbf{y}),\\
\{\Sigma_{I}(\mathbf{x}),\phi^{a}_{J}(\mathbf{y})\}&=&\nu\varepsilon^{0ab}\epsilon_{IJK}e^{K}_{b}\delta^{2} (\mathbf{x}-\mathbf{y}),\\
\{\Phi_{I}(\mathbf{x}),\psi_{J}^{a}(\mathbf{y})\}&=&-\varepsilon^{0ab}\left(\frac{1}{m^{2}}\epsilon_{IJK}f_{b}^{K}+\partial^{\mathbf{x}}_{b}\eta_{IJ}-\epsilon_{IJK}w^{K}_{b}\right)\delta^{2}(\mathbf{x}-\mathbf{y}),\\
\{\Psi_{I}(\mathbf{x}),\psi_{J}^{a}(\mathbf{y})\}&=&\varepsilon^{0ab}\left(\epsilon_{JK}f^{K}_{b}-\left(\nu-m^{2}\right)\left(\partial^{\mathbf{x}}_{b}\eta_{IJ}-\epsilon_{IJK}w^{K}_{b}\right)\right)\delta^{2} (\mathbf{x}-\mathbf{y}),\\
\{\Sigma_{I}(\mathbf{x}),\psi^{a}_{J}(\mathbf{y})\}&=&-m^{2}\varepsilon^{0ab}\epsilon_{IJK}h^{K}_{b}\delta^{2}(\mathbf{x}-\mathbf{y}),\\
\{\Theta_{I}(\mathbf{x}),\psi^{a}_{J}(\mathbf{y})\}&=&m^{2}\varepsilon^{0ab}\epsilon_{IJK}e^{K}_{b}\delta^{2} (\mathbf{x}-\mathbf{y}),\\
\{\Phi_{I}(\mathbf{x}),\sigma^{a}_{J}(\mathbf{y})\}&=&\nu\varepsilon^{0ab}\epsilon_{IJK}e^{K}_{b}\delta^{2}(\mathbf{x}-\mathbf{y}),\\
\{\Psi_{I}(\mathbf{x}),\sigma^{a}_{J}(\mathbf{y})\}&=&-m^{2}\varepsilon^{0ab}\epsilon_{IJK}h^{K}_{b}\delta^{2} (\mathbf{x}-\mathbf{y}),\\
\{\Sigma_{I}(\mathbf{x}),\sigma^{a}_{J}(\mathbf{y})\}&=&\nu\varepsilon^{0ab}\epsilon_{IJK}\left(f^{K}_{b}+2\frac{m^{4}}{\mu}e^{K}_{b}\right)\delta^{2} (\mathbf{x}-\mathbf{y}),\\
\{\Theta_{I}(\mathbf{x}),\sigma^{a}_{J}(\mathbf{y})\}&=&-m^{2}\varepsilon^{0ab}\left(\partial^{\mathbf{x}}_{b}\eta_{IJ}-\epsilon_{IJK}w^{K}_{b}\right)\delta^{2} (\mathbf{x}-\mathbf{y}),\\
\{\Psi_{I}(\mathbf{x}),\theta^{a}_{J}(\mathbf{y})\}&=&-m^{2}\varepsilon^{0ab}\epsilon_{IJK}e^{K}_{b}\delta^{2} (\mathbf{x}-\mathbf{y}),\\
\{\Sigma_{I}(\mathbf{x}),\theta^{a}_{J}(\mathbf{y})\}&=&m^{2}\varepsilon^{0ab}\left(\partial^{\mathbf{x}}_{b}\eta_{IJ}-\epsilon_{IJK}w^{K}_{b}\right)\delta^{2} (\mathbf{x}-\mathbf{y}).\label{Z2}
\end{eqnarray}

\subsection{Poisson brackets between  primary and modified secondary constraints}
\label{Algebra1}
The Poisson brackets of the constraints $\overline{\Phi}_{I}$, $\overline\Sigma_{I}$, $\overline\Theta_{I}$ and $\overline\Sigma_{I}$ with the primary constraints are:
\begin{eqnarray}
\{\overline{\Phi}_{I}(\mathbf{x}),\phi_{J}^{a}(\mathbf{y})\}&=&-\frac{1}{m^{2}}\epsilon_{IJK}\phi^{aK}\delta^{2} (\mathbf{x}-\mathbf{y}),\\
\{\overline{\Phi}_{I}(\mathbf{x}),\psi_{J}^{a}(\mathbf{y})\}&=&\epsilon_{IJK}\phi^{aK}\delta^{2} (\mathbf{x}-\mathbf{y}),\\
\{\overline{\Phi}_{I}(\mathbf{x}),\sigma^{a}_{J}(\mathbf{y})\}&=&-\frac{\nu}{m^{2}}\epsilon_{IJK}\theta^{aK}\delta^{2} (\mathbf{x}-\mathbf{y}),\\
\{\overline\Psi_{I}(\mathbf{x}),\phi_{J}^{a}(\mathbf{y})\}&=&\epsilon_{IJK}\phi^{aK}\delta^{2} (\mathbf{x}-\mathbf{y}),\\
\{\overline\Psi_{I}(\mathbf{x}),\psi_{J}^{a}(\mathbf{y})\}&=&\epsilon_{IJK}\psi^{aK}\delta^{2} (\mathbf{x}-\mathbf{y}),\\
\{\overline\Psi_{I}(\mathbf{x}),\sigma^{a}_{J}(\mathbf{y})\}&=&\epsilon_{IJK}\sigma^{aK}\delta^{2} (\mathbf{x}-\mathbf{y}),\\
\{\overline\Psi_{I}(\mathbf{x}),\theta^{a}_{J}(\mathbf{y})\}&=&\epsilon_{IJK}\theta^{aK}\delta^{2} (\mathbf{x}-\mathbf{y}),\\
\{\overline\Sigma_{I}(\mathbf{x}),\phi^{a}_{J}(\mathbf{y})\}&=&-\frac{\nu}{m^{2}}\epsilon_{IJK}\theta^{aK}\delta^{2} (\mathbf{x}-\mathbf{y}),\\
\{\overline\Sigma_{I}(\mathbf{x}),\psi^{a}_{J}(\mathbf{y})\}&=&\epsilon_{IJK}\sigma^{aK}\delta^{2} (\mathbf{x}-\mathbf{y}),\\
\{\overline\Sigma_{I}(\mathbf{x}),\sigma^{a}_{J}(\mathbf{y})\}&=&m^{2}\epsilon_{IJK}\left(\psi^{aK}-2\frac{\nu}{\mu}\theta^{aK}-\left(\nu-m^{2}\right)\phi^{aK}\right)\delta^{2} (\mathbf{x}-\mathbf{y}),\\
\{\overline\Sigma_{I}(\mathbf{x}),\theta^{a}_{J}(\mathbf{y})\}&=&-\frac{m^{2}}{\nu}\epsilon_{IJK}\left(\psi^{aK}+m^{2}\phi^{aK}\right)\delta^{2} (\mathbf{x}-\mathbf{y}),\\
\{\overline\Theta_{I}(\mathbf{x}),\psi^{a}_{J}(\mathbf{y})\}&=&-\epsilon_{IJK}\theta^{aK}\delta^{2} (\mathbf{x}-\mathbf{y}),\\
\{\overline\Theta_{I}(\mathbf{x}),\sigma^{a}_{J}(\mathbf{y})\}&=&\frac{m^{2}}{\nu}\epsilon_{IJK}\left(m^{2}\phi^{aK}+\psi^{aK}\right)\delta^{2} (\mathbf{x}-\mathbf{y}).
\end{eqnarray}

\subsection{Poisson brackets among secondary constraints}
\label{Algebra2}
The non-trivial Poisson algebra among  modified secondary constraints are:
\begin{eqnarray}
\{\overline{\Phi}_{I}(\mathbf{x}),\overline{\Phi}_{J}(\mathbf{y})\}&=&\left(-\frac{1}{m^{2}}\epsilon_{IJK}\left(\overline{\Phi}^{K}-\frac{\nu}{m^{2}}\epsilon^{KMN}\theta_{aM}e^{a}_{N}\right)+\frac{\nu}{m^{2}}\varepsilon^{0ab}e_{aI}e_{bJ}\right)\delta^{2}(\mathbf{x}-\mathbf{y}),\\
\{\overline{\Phi}_{I}(\mathbf{x}),\overline{\Sigma}_{J}(\mathbf{y})\}&=&\left(\frac{\nu}{m^{2}}\epsilon_{IJK}\overline\Theta^{K}+\epsilon_{I}{^{KN}}\epsilon_{JK}{^{M}}\left(\varepsilon^{0ab}\frac{\nu}{m^{2}}f_{aN}e_{bM}+\left(m^{2}+\nu\right)\phi^{a}_{N}e_{aM}+\frac{\nu}{m^{4}}\theta^{a}_{M}f_{aN}+\psi_{aN}e^{a}_{M}\right)\right)\nonumber\\
&&\times\delta^{2}(\mathbf{x}-\mathbf{y}),\\
\{\overline\Theta_{I}(\mathbf{x}),\overline\Theta_{J}(\mathbf{y})\}&=&\left(\frac{m^{2}}{\nu}\epsilon_{IJK}\epsilon^{KMN}e_{aM}\theta^{a}_{N}-\frac{m^{4}}{\nu}\varepsilon^{0ab}e_{aI}e_{bJ}\right)\delta^{2}(\mathbf{x}-\mathbf{y}),\\
\{\overline\Theta_{I}(\mathbf{x}),\overline\Sigma_{J}(\mathbf{y})\}&=&\left(\frac{m^{2}}{\nu}\epsilon_{IJK}\left(\overline\Psi^{K}+m^{2}\overline\Phi^{K}\right)+\frac{m^{2}}{\nu}\epsilon_{I}{^{KN}}\epsilon_{JK}{^{M}}\left(m^{2}\varepsilon^{0ab}h_{aN}e_{bM}-\theta_{aM}h^{a}_{N}+\sigma_{aN}e^{a}_{M}\right)\right)\delta^{2}(\mathbf{x}-\mathbf{y}),\\
\{\overline\Sigma_{I}(\mathbf{x}),\overline\Sigma_{J}(\mathbf{y})\}&=&\left(m^{2}\epsilon_{IJK}\left(2\frac{\nu}{\mu}\overline\Theta^{K}+\overline\Psi^{K}-\left(\nu-m^{2}\right)\overline\Phi^{K}-\frac{1}{m^{2}}\epsilon^{KMN}\left(\left(\nu-m^{2}\right)f_{aM}\phi^{a}_{N}-f_{aM}\psi^{a}_{N}+\frac{m^{2}}{\nu}h_{aM}\sigma^{a}_{N}\right)\right)\right.\nonumber\\
&&\left.-\frac{\nu}{m^{2}}\varepsilon^{0ab}f_{aI}f_{bJ}-\frac{m^{4}}{\nu}\varepsilon^{0ab}h_{aI}h_{bJ}\right)\delta^{2}(\mathbf{x}-\mathbf{y}).
\end{eqnarray}

\subsection{Poisson brackets between modified secondary and tertiary constraints}
\label{Algebra3}
The non-vanishing Poisson brackets between modified secondary and tertiary constraints are:
\begin{eqnarray}
\{\overline\Phi_{I}(\mathbf{x}),\Xi(\mathbf{y})\}&=&\left(\frac{1}{m^{2}}\Theta_{I}-\frac{1}{m^{2}}\varepsilon^{0ab}\epsilon_{IJK}e_{a}^{J}f_{b}^{K}\right)\delta^{2}(\mathbf{x}-\mathbf{y}),\\
\{\overline\Psi_{I}(\mathbf{x}),\Xi(\mathbf{y})\}&=&0,\\
\{\overline\Theta_{I}(\mathbf{x}),\Xi(\mathbf{y})\}&=&\frac{m^{4}}{\nu}\varepsilon^{0ab}\epsilon_{IJK}e^{J}_{a}e^{K}_{b}\delta^{2}(\mathbf{x}-\mathbf{y}),\\
\{\overline\Sigma_{I}(\mathbf{x}),\Xi(\mathbf{y})\}&=&-\left(\frac{m^{2}}{\nu}\Psi_{I}-\frac{m^{2}}{\nu}\left(\nu-m^{2}\right)\Phi_{I}+\frac{3}{2}m^{2}\left(\nu-m^{2}\right)\varepsilon^{0ab}\epsilon_{IJK}e_{a}^{J}e_{b}^{K}\right.\nonumber\\
&&+\left. 2\frac{m^{4}}{\nu}\varepsilon^{0ab}\epsilon_{IJK}h_{a}^{J}e_{b}^{K}+\frac{1}{2m^{2}}\varepsilon^{0ab}\epsilon_{IJK}f_{a}^{J}f_{b}^{K}\right)\delta^{2}(\mathbf{x}-\mathbf{y}),\\
\{\overline\Phi_{I}(\mathbf{x}),\Upsilon(\mathbf{y})\}&=&-\frac{\nu}{m^{2}}\varepsilon^{0ab}\epsilon_{IJK}e^{J}_{a}e^{K}_{b}\delta^{2}(\mathbf{x}-\mathbf{y}),\\
\{\overline\Psi_{I}(\mathbf{x}),\Upsilon(\mathbf{y})\}&=&0,\\
\{\overline\Sigma_{I}(\mathbf{x}),\Upsilon(\mathbf{y})\}&=&\left(\frac{1}{m^{2}}\Sigma_{I}-\varepsilon^{0ab}\epsilon_{IJK}\left(2\frac{\nu}{m^{2}}f_{a}^{J}e_{b}^{K}+3\frac{\nu}{\mu}m^{2}e_{a}^{J}e_{b}^{K}\right)\right)\delta^{2}(\mathbf{x}-\mathbf{y}),\\
\{\overline\Theta_{I}(\mathbf{x}),\Upsilon(\mathbf{y})\}&=&-\frac{1}{m^{2}}\Theta_{I}\delta^{2}(\mathbf{x}-\mathbf{y}).
\end{eqnarray}

\subsection{Poisson brackets of $\hat{\Sigma}_{I}$ and $\overline{\Psi}$}
\label{Algebra31}
The Poisson brackets of $\hat{\Sigma}_{I}$ with the whole set of constraints have the following form:
\begin{eqnarray}
\{\widehat{\Sigma}_{I}(\mathbf{x}),\phi_{J}^{a}(\mathbf{y})\}&=&\epsilon_{IJK}\left[\frac{m^{2}}{\mu}\phi^{aK}-\frac{\nu}{m^{2}}\theta^{aK}\right]\delta^{2}(\mathbf{x}-\mathbf{y})\approx0,\\
\{\widehat{\Sigma}_{I}(\mathbf{x}),\psi_{J}^{a}(\mathbf{y})\}&=&\epsilon_{IJK}\left[\sigma^{aK}+c_{f}\phi^{aK}-c_{h}\theta^{aK}\right]\delta^{2}(\mathbf{x}-\mathbf{y})\approx0,\\
\{\widehat{\Sigma}_{I}(\mathbf{x}),\sigma_{J}^{a}(\mathbf{y})\}&=&m^{2}\epsilon_{IJK}\left[\frac{1}{2}\left(\frac{m^{2}}{\mu^{2}}+\frac{\left(\nu+m^{2}\right)}{m^{2}}\right)\psi^{aK}+\frac{m^{2}}{2}\left(\frac{m^{2}}{\mu^{2}}-\frac{\left(\nu-m^{2}\right)}{m^{2}}\right)\phi^{aK}-\frac{\nu}{\mu}\theta^{aK}\right]\delta^{2}(\mathbf{x}-\mathbf{y})\nonumber\\
&&\approx0,\\
\{\widehat{\Sigma}_{I}(\mathbf{x}),\theta_{J}^{a}(\mathbf{y})\}&=&-\frac{m^{2}}{\nu}\epsilon_{IJK}\left[\psi^{aK}+m^{2}\phi^{aK}\right]\delta^{2}(\mathbf{x}-\mathbf{y})\approx0,\\
\{\widehat{\Sigma}_{I}(\mathbf{x}),\overline{\Phi}_{J}(\mathbf{y})\}&=&\left[\frac{1}{\mu}\epsilon_{IJK}\left(\nu\overline{\Theta}^{K}+m^{2}\overline{\Phi}^{K}\right)-\epsilon_{IK}{^{M}}\epsilon_{J}{^{KN}}e^{a}_{M}\left(\left(m^{2}+\nu\right)\phi_{aN}+\psi_{aN}-\frac{\nu}{\mu}\theta_{aN}\right)+\frac{\nu}{m^{2}}\eta_{IJ}\Xi\right]\nonumber\\
&&\times\delta^{2}(\mathbf{x}-\mathbf{y})\approx0,\\
\{\widehat{\Sigma}_{I}(\mathbf{x}),\overline{\Theta}_{J}(\mathbf{y})\}&=&\frac{m^{2}}{\nu}\left[\epsilon_{IJK}\left(\overline{\Psi}^{K}+m^{2}\overline{\Phi}^{K}\right)-\epsilon_{IK}{^{M}}\epsilon_{J}{^{KN}}\left(\sigma_{aN}e^{a}_{M}-\theta_{aN}h^{a}_{M}\right)-m^{2}\eta_{IJ}\Upsilon\right]\delta^{2}(\mathbf{x}-\mathbf{y})\approx0,\\
\{\widehat{\Sigma}_{I}(\mathbf{x}),\widehat{\Sigma}_{J}(\mathbf{y})\}&=&\epsilon_{IJK}\left[\left(\nu+\frac{m^{4}}{\mu^{2}}\right)\overline{\Psi}^{K}+\frac{\nu m^{2}}{\mu^{2}}\left(\mu-m^{2}\right)\overline{\Theta}^{K}+\frac{1}{2}\left(\nu-m^{2}\right)\left(1-m^{2}\right)\overline{\Phi}^{K}+2\nu\epsilon^{KMN}\phi_{aM}f^{a}_{N}\right]\nonumber\\
&&\times\delta^{2}(\mathbf{x}-\mathbf{y})\approx0,\\
\{\widehat{\Sigma}_{I}(\mathbf{x}),\overline{\Psi}_{J}(\mathbf{y})\}&=&\epsilon_{IJK}\widehat{\Sigma}^{K}\delta^{2}(\mathbf{x}-\mathbf{y})\approx0,\\
\{\widehat{\Sigma}_{I}(\mathbf{x}),\Xi(\mathbf{y})\}&=&\frac{m^{2}}{\nu}\left[\left(\nu-m^{2}\right)\overline{\Phi}_{I}-\overline{\Psi}_{I}-\frac{\nu}{\mu}\overline{\Theta}_{I}+D_{a}\psi^{a}_{I}-(\nu-m^{2})D_{a}\phi^{a}_{I}-\frac{\nu}{\mu}D_{a}\theta^{a}_{I}-\epsilon_{IJK}\sigma^{J}_{a}e^{aK}\right.\nonumber\\
&&\left.-\frac{m^{2}}{\mu}\epsilon_{IJK}\psi^{J}_{a}e^{aK}-\epsilon_{IJK}\theta^{J}_{a}\left(h^{aK}+\frac{\nu}{m^{2}}\left(\nu-m^{2}\right)e^{aK}\right)-\epsilon_{IJK}\phi^{J}_{a}\left(\frac{m^{4}}{\mu}e^{aK}+\frac{\nu}{m^{2}}f^{aK}\right)\right]\nonumber\\
&&\times\delta^{2}(\mathbf{x}-\mathbf{y})\approx0,\\
\{\widehat{\Sigma}_{I}(\mathbf{x}),\Upsilon(\mathbf{y})\}&=&\frac{1}{m^{2}}\left[\widehat{\Sigma}_{I}+c_{f}\overline{\Phi}_{I}+2c_{h}\overline{\Theta}_{I}-\mathrm{D}_{a}\sigma^{a}_{I}-\frac{\nu}{m^{2}}\epsilon_{IJK}\theta_{a}^{J}f^{aK}-2\frac{\nu}{\mu}m^{2}\epsilon_{IJK}\theta_{a}^{J}e^{aK}-\frac{m^{2}}{\nu}\epsilon_{IJK}\psi_{a}^{J}h^{aK}\right.\nonumber\\
&&+m^{2}\epsilon_{IJK}\psi_{a}^{J}e^{aK}-\frac{m^{4}}{\nu}\epsilon_{IJK}\phi_{a}^{J}h^{aK}-m^{2}\left(\nu-m^{2}\right)\epsilon_{IJK}\phi_{a}^{J}e^{aK}+c_{h}\mathrm{D}_{a}\theta^{a}_{I}+c_{h}\frac{m^{4}}{\nu}\epsilon_{IJK}\phi_{a}^{J}e^{aK}\nonumber\\
&&\left.+c_{h}\frac{m^{2}}{\nu}\epsilon_{IJK}\psi_{a}^{J}e^{aK}\right]\delta^{2}(\mathbf{x}-\mathbf{y})\approx0.
\end{eqnarray}

On the other hand, the Poisson brackets of the constraint $\overline{\Psi}$ with the complete set of constraints are:
\begin{eqnarray}
\{\overline\Psi_{I}(\mathbf{x}),\overline\Psi_{J}(\mathbf{y})\}&=&\epsilon_{IJK}\overline{\Psi}^{K}\delta^{2}(\mathbf{x}-\mathbf{y})\approx0,\\
\{\overline\Psi_{I}(\mathbf{x}),\overline\Phi_{J}(\mathbf{y})\}&=&\epsilon_{IJK}\overline\Phi^{K}\delta^{2}(\mathbf{x}-\mathbf{y})\approx0,\\
\{\overline\Psi_{I}(\mathbf{x}),\widehat\Sigma_{I}(\mathbf{y})\}&=&\epsilon_{IJK}\widehat\Sigma^{K}\delta^{2}(\mathbf{x}-\mathbf{y})\approx0,\\
\{\overline\Psi_{I}(\mathbf{x}),\overline\Theta_{I}(\mathbf{y})\}&=&\epsilon_{IJK}\overline\Theta^{K}\delta^{2}(\mathbf{x}-\mathbf{y})\approx0,\\
\{\overline\Psi_{I}(\mathbf{x}),\phi_{J}^{a}(\mathbf{y})\}&=&\epsilon_{IJK}\phi^{aK}\delta^{2}(\mathbf{x}-\mathbf{y})\approx0,\\
\{\overline\Psi_{I}(\mathbf{x}),\psi_{J}^{a}(\mathbf{y})\}&=&\epsilon_{IJK}\psi^{aK}\delta^{2}(\mathbf{x}-\mathbf{y})\approx0,\\
\{\overline\Psi_{I}(\mathbf{x}),\sigma^{a}_{J}(\mathbf{y})\}&=&\epsilon_{IJK}\sigma^{aK}\delta^{2}(\mathbf{x}-\mathbf{y})\approx0,\\
\{\overline\Psi_{I}(\mathbf{x}),\theta^{a}_{J}(\mathbf{y})\}&=&\epsilon_{IJK}\theta^{aK}\delta^{2}(\mathbf{x}-\mathbf{y})\approx0,\\
\{\overline\Psi_{I}(\mathbf{x}),\Xi(\mathbf{y})\}&=&0,\\
\{\overline\Psi_{I}(\mathbf{x}),\Upsilon(\mathbf{y})\}&=&0.
\end{eqnarray}

 \section{Computation of the inverse Dirac matrix and Dirac Brackets}
 \label{DB}
Using  (\ref{A})-(\ref{D}) and properties of  matrices, one can show that the inverse Dirac matrix turns out to be
\begin{equation}
\label{symplectic}
\left( \boxplus^{AB}\right)^{-1}=
\begin{pmatrix}
 \mathbb{A}^{-1}+\mathbb{A}^{-1}\mathbb{B}\left(\mathbb{D}-\mathbb{C}\mathbb{A}^{-1}\mathbb{B}\right)^{-1}\mathbb{C}\mathbb{A}^{-1} & \quad -\mathbb{A}^{-1}\mathbb{B}\left(\mathbb{D}-\mathbb{C}\mathbb{A}^{-1}\mathbb{B}\right)^{-1}  \\                                                                        
-\left(\mathbb{D}-\mathbb{C}\mathbb{A}^{-1}\mathbb{B}\right)^{-1}\mathbb{C}\mathbb{A}^{-1}   &\quad \left(\mathbb{D}-\mathbb{C}\mathbb{A}^{-1}\mathbb{B}\right)^{-1}
\end{pmatrix}\delta^{2}(\mathbf{x}-\mathbf{y}).
\end{equation}
 with
\begin{equation}
\mathbb{A}^{-1}+\mathbb{A}^{-1}\mathbb{B}\left(\mathbb{D}-\mathbb{C}\mathbb{A}^{-1}\mathbb{B}\right)^{-1}\mathbb{C}\mathbb{A}^{-1} =\epsilon_{0de}
\begin{pmatrix}
 \mathfrak{m}m^{2}     &  \frac{m^{2}}{(\nu-2m^{2})}   &  -\frac{1}{2}\frac{\mu}{m^{2}}\mathfrak{m}    &  \frac{1}{2}\frac{\mu}{m^{2}}c_{h}\mathfrak{m}    \\                                                                        
-\frac{m^{2}}{(\nu-2m^{2})}   &  -\frac{1}{\left(\nu-2m^{2}\right)}      &  -\frac{\mu}{2m^{2}(\nu-2m^{2})}   &   -\frac{\mu}{2m^{2}(\nu-2m^{2})}c_{h}  \\
    \frac{\mu}{2m^{2}}\mathfrak{m}     &  \frac{\mu}{2m^{2}(\nu-2m^{2})}    & \frac{1}{2}\frac{m^{2}}{c_{f}^{2}}\mathfrak{m}  & -\frac{1}{m^{2}}+m^{2}\frac{c_{h}}{c_{f}^{2}}\mathfrak{m} \\
-\frac{\mu}{2m^{2}}c_{h}\mathfrak{m}  &   \frac{\mu}{2m^{2}(\nu-2m^{2})}c_{h}  &\frac{1}{m^{2}}-m^{2}\frac{c_{h}}{c^{2}_{f}}\mathfrak{m} & m^{2}\frac{c_{h}^{2}}{c_{f}^{2}} 	
\end{pmatrix}\eta^{MN},
\end{equation}

\begin{equation}
-\mathbb{A}^{-1}\mathbb{B}\left(\mathbb{D}-\mathbb{C}\mathbb{A}^{-1}\mathbb{B}\right)^{-1}  =\frac{3}{2\mathbf{e}}
\begin{pmatrix}
0     & 0   &  0    &  -\frac{1}{2}\mu\mathfrak{m}    \\                                                                        
0   & 0   &   0&-\frac{1}{2}\frac{\mu}{(\nu-2m^{2})}  \\
    0&  0   & \frac{1}{c_{f}}  & -\frac{1}{2}\frac{\mu}{c_{f}}\mathfrak{m} \\
0  &   0  &\frac{c_{h}}{c_{f}} & -\frac{3}{2}\left(1+\mu\frac{c_{h}}{c_{f}}\mathfrak{m}\right)
\end{pmatrix}\epsilon^{NIL}e_{bI}e_{0L},
\end{equation}

\begin{equation}
-\left(\mathbb{D}-\mathbb{C}\mathbb{A}^{-1}\mathbb{B}\right)^{-1}\mathbb{C}\mathbb{A}^{-1}  =\frac{3}{2\mathbf{e}}
\begin{pmatrix}
0     & 0   &  0    &  0    \\                                                                        
0   & 0   &   0&0 \\
    0&  0   & \frac{1}{c_{f}}  & -\frac{c_{h}}{c_{f}} \\
\frac{1}{2}\mu\mathfrak{m}  &   \frac{1}{2}\frac{\mu}{(\nu-2m^{2})}  &\frac{1}{2}\frac{\mu}{c_{f}}\mathfrak{m} & \frac{3}{2}\left(1-\mu\frac{c_{h}}{c_{f}}\mathfrak{m}\right)
\end{pmatrix}\epsilon^{NIL}e_{bI}e_{0L},
\end{equation}

\begin{equation}
\left(\mathbb{D}-\mathbb{C}\mathbb{A}^{-1}\mathbb{B}\right)^{-1}  =\frac{3}{2\mathbf{e}}
\begin{pmatrix}
\frac{m^{2}}{\nu}     & 0   &  0    &  0    \\                                                                        
0   & -\frac{\nu}{m^{4}}   &   0&0  \\
    0&  0   & 0  & \frac{1}{2}\frac{\mu}{m^{2}}\mathfrak{m} \\
0  &   0  &-\frac{1}{2}\frac{\mu}{m^{2}} & \frac{\mu^{2}}{m^{2}}\mathfrak{m}
\end{pmatrix}\epsilon^{NIL}e_{0L}.
\end{equation}
Here we have abbreviated $\mathfrak{m}=\frac{\nu-m^{2}}{\nu-2m^{2}}$, ${\bf e}=\text{det}|e_{\alpha}^{I}|\neq0$

Using Eq. (\ref{FormDB}), we thus find that the Dirac brackets among all the phase space variables are given by:
 \begin{eqnarray}
\{e_{I}^{a}(\mathbf{x}),\tilde{\Pi}_{b}^{J}(\mathbf{y})\}_{\text{D}}&=&-\frac{1}{2}\mu\frac{c_{h}}{c_{f}}\mathfrak{m}\delta^{a}_{b}\delta^{J}_{I}\delta^{2}(\mathbf{x}-\mathbf{y}),\\
\{e_{I}^{a}(\mathbf{x}),e^{b}_{J}(\mathbf{y})\}_{\text{D}}&=&\frac{1}{2}\frac{m^{2}}{c_{f}^{2}}\mathfrak{m}\varepsilon^{0ab}\eta_{IJ}\delta^{2}(\mathbf{x}-\mathbf{y})\\
\{e_{I}^{a}(\mathbf{x}),f^{b}_{J}(\mathbf{y})\}_{\text{D}}&=&\frac{1}{2}\frac{m^{2}}{c_{f}}\mathfrak{m}\varepsilon^{0ab}\eta_{IJ}\delta^{2}(\mathbf{x}-\mathbf{y}),\\
\{e_{I}^{a}(\mathbf{x}),h^{b}_{J}(\mathbf{y})\}_{\text{D}}&=&\frac{1}{2}m^{2}\frac{c_{h}}{c_{f}^{2}}\mathfrak{m}\varepsilon^{0ab}\eta_{IJ}\delta^{2}(\mathbf{x}-\mathbf{y}),\\
\{e_{I}^{a}(\mathbf{x}),w^{b}_{J}(\mathbf{y})\}_{\text{D}}&=&-\frac{1}{2}\frac{m^{2}}{c_{f}}\frac{1}{\left(\nu-2m^{2}\right)}\varepsilon^{0ab}\eta_{IJ}\delta^{2}(\mathbf{x}-\mathbf{y})\\
\{e_{I}^{a}(\mathbf{x}),\Pi_{b}^{J}(\mathbf{y})\}_{\text{D}}&=&\frac{3}{4}\frac{1}{c_{f}}\mathfrak{m}\delta^{a}_{b}\delta_{I}^{J}\delta^{2}(\mathbf{x}-\mathbf{y}),\\
\{e_{I}^{a}(\mathbf{x}),\tilde{\pi}_{b}^{J}(\mathbf{y})\}_{\text{D}}&=&-\frac{1}{2}\frac{\mu}{c_{f}}\mathfrak{m}\delta^{a}_{b}\delta_{I}^{J}\delta^{2}(\mathbf{x}-\mathbf{y}),\\
\{e_{I}^{a}(\mathbf{x}),\pi_{b}^{J}(\mathbf{y})\}_{\text{D}}&=&\frac{1}{4}\frac{m^{2}}{c_{f}}\mathfrak{m}\delta^{a}_{b}\delta_{I}^{J}\delta^{2}(\mathbf{x}-\mathbf{y}),\\
\{w_{I}^{a}(\mathbf{x}),\pi_{b}^{J}(\mathbf{y})\}_{\text{D}}&=&\left(1+\frac{1}{2}\mathfrak{m}\right)\delta^{a}_{b}\delta_{I}^{J}\delta^{2}(\mathbf{x}-\mathbf{y}),\\
\{w_{I}^{a}(\mathbf{x}),\Pi_{b}^{J}(\mathbf{y})\}_{\text{D}}&=&\frac{1}{2}\mathfrak{m}\delta^{a}_{b}\delta_{I}^{J}\delta^{2}(\mathbf{x}-\mathbf{y}),\\
\{w_{I}^{a}(\mathbf{x}),\tilde{\pi}_{b}^{J}(\mathbf{y})\}_{\text{D}}&=&0,\\
\{w_{I}^{a}(\mathbf{x}),\tilde{\Pi}_{b}^{J}(\mathbf{y})\}_{\text{D}}&=&\frac{1}{2}\frac{\mu}{\left(\nu-2m^{2}\right)}c_{h}\delta^{a}_{b}\delta_{I}^{J}\delta^{2}(\mathbf{x}-\mathbf{y}),\\
\{w_{I}^{a}(\mathbf{x}),f^{b}_{J}(\mathbf{y})\}_{\text{D}}&=&\frac{m^{2}}{\left(\nu-2m^{2}\right)}\delta^{a}_{b}\eta_{IJ}\delta^{2}(\mathbf{x}-\mathbf{y}),\\
\{w_{I}^{a}(\mathbf{x}),h^{b}_{J}(\mathbf{y})\}_{\text{D}}&=&\frac{1}{2}\frac{\mu}{m^{2}}\frac{1}{\left(\nu-2m^{2}\right)}c_{h}\varepsilon^{0ab}\eta_{IJ}\delta^{2}(\mathbf{x}-\mathbf{y}),\\
\{w_{I}^{a}(\mathbf{x}),w^{b}_{J}(\mathbf{y})\}_{\text{D}}&=&-\frac{1}{\left(\nu-2m^{2}\right)}\varepsilon^{0ab}\eta_{IJ}\delta^{2}(\mathbf{x}-\mathbf{y}),\\
\{h_{I}^{a}(\mathbf{x}),\tilde{\pi}_{b}^{J}(\mathbf{y})\}_{\text{D}}&=&-\frac{1}{2}\mu\frac{c_{h}}{c_{f}}\mathfrak{m}\delta^{a}_{b}\delta_{I}^{J}\delta^{2}(\mathbf{x}-\mathbf{y}),\\
\{h_{I}^{a}(\mathbf{x}),\tilde{\Pi}_{b}^{J}(\mathbf{y})\}_{\text{D}}&=&\mu\frac{c_{h}^{2}}{c_{f}}\mathfrak{m}\delta^{a}_{b}\delta^{J}_{I}\delta^{2}(\mathbf{x}-\mathbf{y}),\\
\{h_{I}^{a}(\mathbf{x}),\Pi_{b}^{J}(\mathbf{y})\}_{\text{D}}&=&\frac{3}{4}\frac{c_{h}}{c_{f}}\mathfrak{m}\delta^{a}_{b}\delta_{I}^{J}\delta^{2}(\mathbf{x}-\mathbf{y}),\\
\{h_{I}^{a}(\mathbf{x}),\pi_{b}^{J}(\mathbf{y})\}_{\text{D}}&=&\frac{1}{4}m^{2}\frac{c_{h}}{c_{f}}\mathfrak{m}\delta^{a}_{b}\delta_{I}^{J}\delta^{2}(\mathbf{x}-\mathbf{y}),\\
\{f_{I}^{a}(\mathbf{x}),h^{b}_{J}(\mathbf{y})\}_{\text{D}}&=&-\frac{1}{2}m^{2}\frac{c_{h}}{c_{f}}\mathfrak{m}\varepsilon^{0ab}\eta_{IJ}\delta^{2}(\mathbf{x}-\mathbf{y}),\\
\{h_{I}^{a}(\mathbf{x}),h^{b}_{J}(\mathbf{y})\}_{\text{D}}&=&\frac{1}{2}m^{2}\frac{c_{h}^{2}}{c_{f}^{2}}\mathfrak{m}\varepsilon^{0ab}\eta_{IJ}\delta^{2}(\mathbf{x}-\mathbf{y}),\\
\{f_{I}^{a}(\mathbf{x}),\Pi_{b}^{J}(\mathbf{y})\}_{\text{D}}&=&\frac{3}{2}\mathfrak{m}\delta^{a}_{b}\delta_{I}^{J}\delta^{2}(\mathbf{x}-\mathbf{y}),\\
\{f_{I}^{a}(\mathbf{x}),f^{b}_{J}(\mathbf{y})\}_{\text{D}}&=&\frac{1}{2}m^{2}\mathfrak{m}\varepsilon^{0ab}\eta_{IJ}\delta^{2}(\mathbf{x}-\mathbf{y}).
\end{eqnarray}

\end{document}